\documentclass[11pt]{article}
\usepackage{graphicx}
\usepackage{amsmath}
\usepackage{amsfonts}
\usepackage{amssymb}
\usepackage{hyperref}
\usepackage{cite}
\usepackage{geometry}
\usepackage{multicol}
\usepackage{booktabs}

\title{CP Violation and Flavour-Violating Di-Higgs Couplings in the Randall-Sundrum Model}
\author{Gayatri Ghosh \\ Department Of Physics, Cachar College \\ Email: gayatrighsh@gmail.com.com}
\date{\today}

\begin{document}

\maketitle

\begin{abstract}
 The Randall-Sundrum (RS) model offers a compelling framework to address the hierarchy problem and provides new sources of CP violation beyond the Standard Model (SM). The motivation for studying CP violation in the RS model arises from the insufficiency of CP-violating phases in the SM to account for the observed matter-antimatter asymmetry in the universe. In this work, we explore CP violation through flavour-violating di-Higgs couplings, which emerge due to the localization of bulk fermions and the Higgs near the TeV brane. The analysis focuses on the role of these couplings in di-Higgs production and decay processes, leading to enhanced CP-violating effects. Numerical simulations show that the predicted CP-violating observables are within experimental bounds and could be tested in future collider experiments. The study concludes that flavour-violating di-Higgs couplings in the RS model offer a promising avenue for discovering new sources of CP violation, with significant implications for both collider physics and the understanding of the matter-antimatter asymmetry.
\end{abstract}

\section{Introduction}
The discovery of the Higgs boson at the Large Hadron Collider (LHC) has opened new avenues for exploring the dynamics of electroweak symmetry breaking. Theoretical frameworks beyond the Standard Model (BSM), such as the Randall-Sundrum model, offer compelling explanations for the Higgs mechanism through the presence of extra dimensions. This paper focuses on CP violation and flavor-violating di-Higgs couplings, highlighting their significance in understanding the interplay between flavor physics and the dynamics of the Higgs sector.

The Standard Model (SM) of particle physics has been remarkably successful in explaining many phenomena, yet it leaves several questions unresolved. One of the most significant issues is the \textit{hierarchy problem}, which concerns the large difference between the electroweak scale ($v \approx 246\,\text{GeV}$) and the Planck scale ($M_{\text{Pl}} \approx 10^{19}\,\text{GeV}$). This vast discrepancy suggests that the Higgs mass requires extreme fine-tuning in the absence of new physics. The Randall-Sundrum (RS) model provides an elegant solution to the hierarchy problem by introducing an extra spatial dimension, compactified on a \textit{warped geometry}.

The RS model postulates a five-dimensional Anti-de Sitter ($AdS_5$) space where the metric is given by:
\begin{equation}
ds^2 = e^{-2kr_c|\phi|} \eta_{\mu\nu} dx^\mu dx^\nu - r_c^2 d\phi^2,
\end{equation}
where $k$ is the $AdS_5$ curvature scale, $r_c$ is the compactification radius, and $\phi$ is the angular coordinate of the extra dimension. The Higgs field is localized on the TeV brane at $\phi = \pi$. The warp factor, $e^{-2kr_c|\phi|}$, naturally generates an exponential hierarchy between the weak and Planck scales, such that:
\begin{equation}
M_{\text{Pl}} e^{-kr_c \pi} \sim \mathcal{O}(\text{TeV}),
\end{equation}
solving the hierarchy problem without fine-tuning.

In the SM, CP violation is introduced via the complex phase in the Cabibbo-Kobayashi-Maskawa (CKM) matrix. While this explains the CP violation observed in the quark sector, it is insufficient to account for the observed matter-antimatter asymmetry of the universe. Beyond the SM, theories like Supersymmetry (SUSY), Composite Higgs models, and extra-dimensional models like RS predict new sources of CP violation.

In the RS model, CP violation can occur due to the localization of fermions in the extra-dimensional bulk. The Yukawa couplings, which generate masses for fermions, become position-dependent in the extra dimension. The overlap of fermion wavefunctions with the Higgs localized on the TeV brane introduces flavor-dependent and CP-violating effects. The 4D effective Yukawa couplings can be written as:
\begin{equation}
\lambda_{ij}^{4D} = \int_0^{\pi} d\phi \, e^{-3kr_c|\phi|} f_i(\phi) f_j(\phi) h(\phi),
\end{equation}
where $f_i(\phi)$ and $f_j(\phi)$ are the fermion wavefunctions in the extra dimension, and $h(\phi)$ is the Higgs profile. These integrals result in off-diagonal Yukawa couplings, leading to flavor-changing and CP-violating effects.

CP violation beyond the SM has been extensively studied in various frameworks:

\begin{itemize}
    \item Supersymmetry (SUSY): In SUSY models, CP-violating phases arise from the soft SUSY-breaking terms. The minimal supersymmetric standard model (MSSM) includes phases in the gaugino masses, the $\mu$-parameter, and the trilinear $A$-terms, which contribute to electric dipole moments (EDMs) and flavor-violating processes.
    \item Composite Higgs Models: In composite Higgs scenarios, where the Higgs is a pseudo-Goldstone boson of a new strong interaction, CP violation can arise through interactions between the Higgs and composite fermions.
    \item Leptoquark Models: Leptoquarks offer another framework where CP violation can occur, particularly in flavor-changing neutral currents (FCNCs) and processes involving meson mixing and rare decays.
\end{itemize}

In the RS model, CP-violating phases can be introduced via complex phases in the 5D Yukawa couplings. These phases generate new CP-violating observables, such as EDMs and flavor-violating Higgs decays, which could be probed in future collider experiments.

The RS model provides a natural solution to the hierarchy problem and offers new sources of CP violation through flavor-violating interactions. By studying the flavor structure of the model and its impact on CP-violating observables, we can explore new physics beyond the SM. The RS model, along with other frameworks like SUSY and Composite Higgs, remains a promising candidate for addressing both the hierarchy problem and the matter-antimatter asymmetry.
\par 
The Standard Model (SM) has been highly successful in describing fundamental particle interactions, but it fails to explain certain crucial phenomena, such as the hierarchy problem and the observed matter-antimatter asymmetry in the universe. The hierarchy problem is the large disparity between the electroweak scale ($v \approx 246\,\text{GeV}$) and the Planck scale ($M_{\text{Pl}} \approx 10^{19}\,\text{GeV}$). The Randall-Sundrum (RS) model provides a natural solution to the hierarchy problem by introducing a warped extra dimension, compactified on an \textit{AdS}$_5$ background.

Beyond solving the hierarchy problem, the RS model also opens up new possibilities for studying CP violation. In the SM, CP violation is introduced through the Cabibbo-Kobayashi-Maskawa (CKM) matrix, but it is insufficient to account for the magnitude of matter-antimatter asymmetry observed in the universe. This motivates the exploration of additional sources of CP violation, which naturally arise in the RS framework due to the localization of fermions in the extra-dimensional bulk.

The primary goal of this paper is to investigate the role of CP violation in the RS model, focusing on flavour-violating di-Higgs couplings \cite{a}. These couplings emerge as a consequence of the position-dependent fermion wavefunctions interacting with the Higgs boson localized on the TeV brane. Such interactions introduce new sources of CP violation, which can be probed through di-Higgs production and decay processes at colliders, such as the Large Hadron Collider (LHC).

In this study, we perform detailed numerical simulations to explore the parameter space of the RS model, incorporating existing experimental constraints and predicting new CP-violating observables. Our results provide a comprehensive analysis of the phenomenological implications of flavour-violating di-Higgs couplings, highlighting the potential for future experimental verification.

The production of multiple Higgs bosons in high-energy collisions has been one of the central objectives of particle physics in the search for beyond-the-Standard-Model (BSM) phenomena. Di-Higgs production, in particular, provides a unique window into probing the properties of the Higgs potential and testing models such as the Randall-Sundrum (RS) model with extra dimensions. This study focuses on the enhancement of di-Higgs production \cite{b1} through the inclusion of Kaluza-Klein (KK) gravitons, CP-violating interactions, and flavor-violating terms within the RS framework.

In the RS model, the hierarchy problem is addressed by introducing an extra spatial dimension compactified on a \( S^1/Z_2 \) orbifold, leading to a rich spectrum of KK excitations. These KK modes significantly modify the Higgs production cross-sections, particularly in processes such as gluon-gluon fusion. Furthermore, the mixing between the radion and the Higgs boson introduces additional contributions, which, together with CP violation, give rise to distinct experimental signatures.

The inclusion of higher-dimensional operators and CP-violating interactions in the RS model is essential to account for possible new physics contributions beyond the Standard Model. The detailed theoretical derivations for these operators and their corresponding interactions with the Higgs and KK graviton fields are provided in Appendix A. These include the effective 4D Lagrangian for the radion, the KK gravitons, and the flavor-violating couplings that arise due to fermion localization in the bulk. The role of these terms in modifying the di-Higgs production cross-section is explored numerically in the context of parton-level simulations.

In addition, flavor violation plays a crucial role in RS models with bulk fermions, leading to observable effects in both collider experiments and precision tests. The interplay between flavor violation and CP-violating interactions is carefully examined in Appendix B, where we present a detailed calculation of the cross-sections for di-Higgs production, as well as the corrections introduced by the radion-Higgs mixing. 

The main results of this paper are obtained by performing a Monte Carlo simulation of parton-level events, followed by fast detector simulation. The key computational steps, including event generation, showering, hadronization, and detector response, are outlined in Appendix C. Our findings demonstrate that the RS model, with its distinctive radion-Higgs interactions and CP-violating terms, enhances di-Higgs production significantly, particularly at high center-of-mass energies.

The structure of the paper is as follows: in Section 2, we review the basics of the RS model and describe the modifications introduced by CP violation and flavor violation. In Section 3, we provide an overview of the Monte Carlo simulation framework used to evaluate the cross-sections for di-Higgs production. Section 4 presents the numerical results, with a discussion on the experimental implications. Finally, we conclude in Section 5, where we summarize our findings and outline future directions for research.

\section{The Randall-Sundrum Model and Its Higgs Sector}

\subsection{Randall-Sundrum Model}

The Randall-Sundrum (RS) model introduces a non-factorizable warped geometry with a 5-dimensional metric of the form:
\begin{equation}
ds^2 = e^{-2 k r_c |y|} \eta_{\mu\nu} dx^\mu dx^\nu - r_c^2 dy^2,
\end{equation}
where \( k \) is the curvature scale of the extra dimension, \( r_c \) is the compactification radius, and \( y \in [-\pi, \pi] \) represents the coordinate along the extra dimension. The warp factor \( e^{-2 k r_c |y|} \) exponentially suppresses energy scales as one moves from the Planck brane (\( y = 0 \)) to the TeV brane (\( y = \pi \)), thus solving the hierarchy problem by reducing the Planck scale to the TeV scale at the brane located at \( y = \pi \).

The 4-dimensional effective Planck scale, \( M_{\text{Pl}} \), is related to the 5-dimensional Planck scale, \( M_5 \), by integrating over the extra dimension:
\begin{equation}
M_{\text{Pl}}^2 = \frac{M_5^3}{k} \left( 1 - e^{-2 k r_c \pi} \right),
\end{equation}
which for large \( k r_c \) gives the desired suppression of the Planck scale to the electroweak scale.

\subsection{Higgs Localization}

In the RS model, the Higgs field can either be localized on the TeV brane or propagate in the bulk. When the Higgs is confined to the TeV brane, its vacuum expectation value (VEV) is naturally suppressed by the warp factor, ensuring the hierarchy between the electroweak and Planck scales.

\paragraph{TeV Brane Higgs.}
For a Higgs field \( H \) localized on the TeV brane, the Higgs VEV \( v \) appears naturally as:
\begin{equation}
v_{\text{eff}} = e^{-k r_c \pi} v_0,
\end{equation}
where \( v_0 \) is the VEV in the absence of warping. This exponential suppression allows \( v_{\text{eff}} \sim 246\,\text{GeV} \), despite \( v_0 \sim M_{\text{Pl}} \).

\paragraph{Bulk Higgs.}
Alternatively, the Higgs field can be allowed to propagate in the bulk. The effective 4D profile of the Higgs field \( H(y) \) in this case depends on its localization parameter \( \nu \), where:
\begin{equation}
H(y) \sim e^{(2 - \nu) k r_c |y|}.
\end{equation}
For \( \nu > 2 \), the Higgs is localized closer to the Planck brane, while for \( \nu < 2 \), it is localized towards the TeV brane. A bulk Higgs offers a more flexible framework for addressing flavor hierarchies through fermion profiles.

\subsection{Higgs Potential}

The Higgs potential on the TeV brane, in the RS setup, is typically given by:
\begin{equation}
V(H) = \lambda_H \left( H^\dagger H - v^2 \right)^2 + \left( \mu_H^2 H^2 + \lambda_{HA} HA + \text{h.c.} \right),
\end{equation}
where \( \lambda_H \) is the Higgs self-coupling, \( v \) is the Higgs VEV, and \( \mu_H \) and \( \lambda_{HA} \) are coupling constants that can be complex, potentially introducing CP-violating terms.

CP violation can arise from complex phases in these couplings. Let \( \lambda_{HA} \) and \( \mu_H^2 \) be complex:
\begin{equation}
\lambda_{HA} = |\lambda_{HA}| e^{i \delta_{\lambda}}, \quad \mu_H^2 = |\mu_H^2| e^{i \delta_{\mu}}.
\end{equation}
The presence of these complex phases \( \delta_{\lambda} \) and \( \delta_{\mu} \) introduces explicit CP-violating interactions in the Higgs sector. These CP-violating terms can manifest in processes involving the Higgs, such as di-Higgs production and rare decays, which can be tested at high-energy colliders.

The RS model's Higgs sector provides a rich framework for exploring both the hierarchy problem and new sources of CP violation. By introducing complex couplings, the model predicts measurable CP-violating observables, which could be probed in future collider experiments like the LHC.

\begin{equation}
V(H) = \lambda_H \left( H^\dagger H - v^2 \right)^2 + \left( |\mu_H^2| e^{i \delta_{\mu}} H^2 + |\lambda_{HA}| e^{i \delta_{\lambda}} HA + \text{h.c.} \right).
\end{equation}

These complex terms can lead to observable CP-violating effects in di-Higgs interactions and flavour-violating processes. Such effects can provide crucial tests of the RS model and its potential for explaining CP violation beyond the Standard Model.

\begin{figure}[h!]
    \centering
    \includegraphics[width=0.7\textwidth]{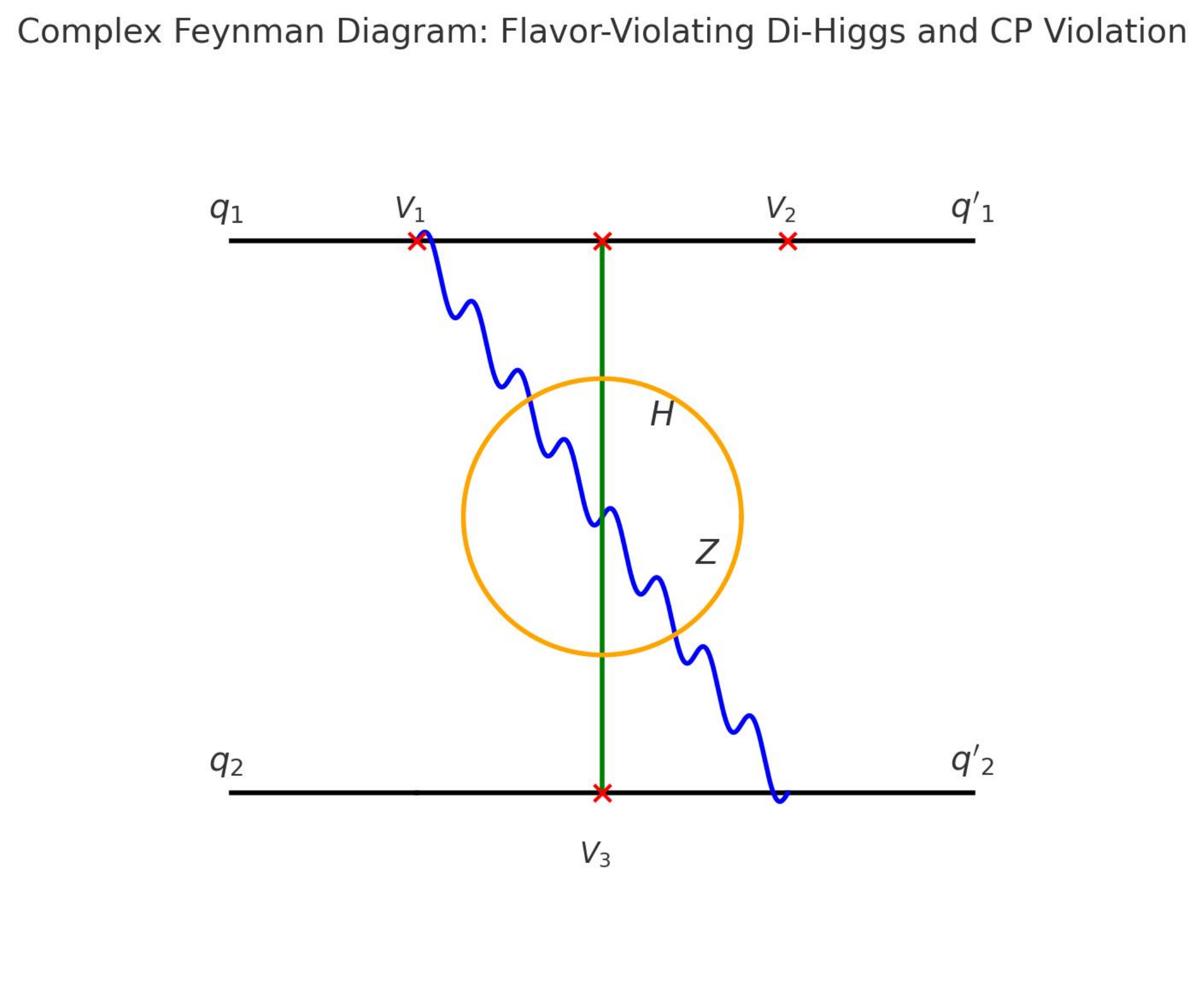} 
    \caption{Feynman diagram showing processes like di-Higgs production with flavor-changing neutral currents (FCNC) and CP violation. The diagram includes multiple fermion lines, wavy lines for Higgs and gauge bosons, as well as a loop representing higher-order quantum corrections.}
    \label{fig:complex_feynman_diagram}
\end{figure}

In Fig. 1 the Feynman diagram presented illustrates a process involving flavor-changing neutral currents (FCNC) and CP violation through di-Higgs production. The diagram consists of the following elements:

\begin{itemize}

    \item \textbf{External Fermion Lines:} Two sets of external fermions are present: 
    \[
    q_1 \rightarrow q'_1 \quad \text{and} \quad q_2 \rightarrow q'_2
    \]
    These lines represent initial and final-state quarks. The fermion lines are straight and denoted by $q_1$, $q'_1$, $q_2$, and $q'_2$. These transitions involve flavor-changing processes where the flavor of the quarks can change during the interaction, characteristic of FCNC \cite{b2}.

    \item \textbf{Internal Fermion Propagators:} The diagram shows internal fermion lines represented by $V_1$, $V_2$, and $V_3$. These vertices correspond to places where gauge boson exchange or loop interactions take place, mediating the transitions between the fermions. In the center of the diagram, wavy lines represent the propagation of gauge bosons (e.g., $Z$) and Higgs bosons ($H$). These bosons mediate the interaction between the fermions and are crucial in allowing both FCNC and CP violation. The Higgs boson is responsible for generating mass and enabling CP violation through complex couplings. The $Z$ boson exchange ensures neutral current processes. There is a loop shown between the fermion propagators and the wavy lines. This loop represents higher-order quantum corrections, contributing to the CP-violating phase through interference between various interaction terms. In such processes, the loop can introduce complex phases in the couplings, allowing CP violation to manifest. The exchange of the $Z$ boson and the Higgs boson permits flavor-changing neutral current processes, where the quark flavor changes without altering the charge. These processes are heavily suppressed in the Standard Model, making them ideal for investigating new physics, including CP violation. Fig. 1 captures the essential dynamics of di-Higgs production with FCNC and CP violation, where quantum corrections and boson exchanges facilitate interactions between different quark flavors while allowing for CP-violating effects via the complex couplings of the Higgs and gauge bosons.

\end{itemize}

The exploration of flavor physics and CP violation is crucial for understanding the limitations of the Standard Model (SM) and probing for new physics. The Randall-Sundrum (RS) model, an extra-dimensional theory, offers a compelling framework for flavor violation through the localization of fermions in the extra dimension. This paper discusses how flavor-violating di-Higgs couplings, particularly involving the tau and muon flavors characterized by effective coupling constants such as \(C_{\mu\tau}\), can give rise to CP violation, comparing the implications with other Beyond Standard Model (BSM) theories.

The exploration of flavor physics and CP violation is crucial for understanding the limitations of the Standard Model (SM) and probing for new physics. The Randall-Sundrum (RS) model, an extra-dimensional theory, offers a compelling framework for flavor violation through the localization of fermions in the extra dimension. This localization leads to flavor-violating di-Higgs couplings, particularly involving the tau and muon flavors, characterized by effective coupling constants such as \(C_{\mu\tau}\). 

\section{Framework of the Randall-Sundrum Model}
The RS model is based on a five-dimensional spacetime with a warped metric:

\begin{equation}
ds^2 = e^{-2k |y|} \eta_{\mu\nu} dx^\mu dx^\nu - dy^2,
\end{equation}

where \(k\) is the curvature scale, \(y\) is the extra-dimensional coordinate, and \(\eta_{\mu\nu}\) is the Minkowski metric. The SM fermions are assigned to different locations in the extra dimension, leading to distinct wavefunctions and couplings.

\subsection{Flavor-Violating Couplings}
In the RS model, the effective Yukawa couplings can be expressed as:

\begin{equation}
\mathcal{L}_{\text{Yukawa}} = - \sum_{i,j} y_{ij} \bar{Q}_i H Q_j + \text{h.c.}
\end{equation}

In the context of flavor violation, we include additional terms representing flavor-violating interactions:

\begin{equation}
\mathcal{L}_{\text{FV}} = - \sum_{i \neq j} C_{ij} \bar{Q}_i H Q_j + \text{h.c.}
\end{equation}

where \(C_{ij}\) are flavor-violating couplings, for instance, \(C_{\mu\tau}\) and others that may couple various flavors, including those of the muon and tau leptons.

\subsection{Di-Higgs Couplings}
The effective potential for di-Higgs interactions can be generalized to include flavor-violating terms:

\begin{equation}
V(h, h) = \lambda_{hh} h^2 + \lambda_{h \text{f}} h \bar{f} f + \lambda_{hh \text{f}} h^2 \bar{f} f + \sum_{i,j} C_{ij} h^2 \bar{f}_i f_j,
\end{equation}

where \(\lambda_{hh}\), \(\lambda_{h \text{f}}\), and \(\lambda_{hh \text{f}}\) are the Higgs self-couplings and couplings to fermions.

\subsection{Flavor Violating Di-Higgs Couplings}
The flavor-violating di-Higgs couplings can be represented as follows:

\begin{equation}
\mathcal{L}_{\text{FV-DiHiggs}} = C_{\mu\tau} h^2 \bar{\mu} \tau + C_{b\tau} h^2 \bar{b} \tau + C_{b\mu} h^2 \bar{b} \mu + \ldots,
\end{equation}

where \(C_{\mu\tau}\), \(C_{b\tau}\), and \(C_{b\mu}\) represent the flavor-violating couplings between the di-Higgs and fermion pairs.

\section{Mechanisms of CP Violation}
\subsection{CP Violation from Flavor-Violating Couplings}
The flavor-violating couplings \(C_{ij}\) can introduce complex phases into the interactions, leading to CP-violating phenomena. The effective Lagrangian can include:

\begin{equation}
\mathcal{L}_{\text{eff}} = \sum_{i,j} \left( C_{ij} e^{i \phi_{ij}} \bar{Q}_i H Q_j + \text{h.c.} \right),
\end{equation}

where \(\phi_{ij}\) are the complex phases associated with the couplings. The Jarlskog invariant \(J\) quantifies CP violation, given by:

\begin{equation}
J = \text{Im} \left( C_{ij} C_{kl}^* C_{il}^* C_{kj} \right).
\end{equation}

\subsection{Phenomenological Implications}
The interplay between flavor violation and CP violation has important implications for experimental observables, particularly in B-meson decays. The contributions of the flavor-violating di-Higgs couplings to processes like \(b \to s \gamma\) and \(B_s \to \mu^+\mu^-\) can be described by the effective Hamiltonian:

\begin{equation}
\mathcal{H}_{\text{eff}} = \frac{G_F}{\sqrt{2}} \left( V_{tb} V_{ts}^* \left( C_7^{\text{eff}} O_7 + C_9 O_9 + C_{10} O_{10} \right) + \ldots \right),
\end{equation}

where \(O_i\) are the relevant operators and \(C_i\) include contributions from the flavor-violating di-Higgs interactions.

\section{Calculations of Flavor-Violating Couplings in RS Model}
In the context of the RS model, the flavor-violating couplings can be expressed in terms of parameters such as the fermion masses and their localization in the extra dimension. We can compute \(C_{ij}\) as follows:

\begin{equation}
C_{ij} = \int dy \, \psi_i(y) \psi_j(y) e^{-k|y|},
\end{equation}

where \(\psi_i(y)\) and \(\psi_j(y)\) are the wavefunctions of the fermions in the extra dimension. The parameters can be varied to observe the effects of localization on \(C_{ij}\).

Assuming the fermions are localized at different points in the extra dimension, we can model the wavefunctions as:

\begin{equation}
\psi_i(y) = A_i e^{-m_i |y|},
\end{equation}

where \(m_i\) is the mass of the fermion and \(A_i\) is the normalization constant. Thus, we can calculate \(C_{\mu\tau}\):

\begin{equation}
C_{\mu\tau} = A_\mu A_\tau \int dy \, e^{-(m_\mu + m_\tau)|y|} e^{-k|y|}.
\end{equation}

This integral can be computed for specific values of \(m_\mu\), \(m_\tau\), and \(k\).

\section{Comparison with Other BSM Theories}
\subsection{Minimal Supersymmetric Standard Model (MSSM)}
In the MSSM, the flavor-violating contributions primarily arise from complex phases in the soft breaking terms. The flavor-violating couplings in the MSSM can lead to similar processes but are often constrained by flavor-changing neutral currents. In contrast, the RS model’s flavor violation is more directly tied to the geometry of the extra dimension, allowing for potentially observable di-Higgs processes that escape such constraints.

\subsection{Composite Higgs Models}
In composite Higgs models, flavor violation can arise from the dynamics of the strong sector. While such models can lead to non-trivial flavor structure, the RS model's warped geometry provides unique mechanisms for flavor violation through localization effects. This distinction may lead to different signatures in Higgs production and decay, offering a path for experimental differentiation.

The Randall-Sundrum model provides a robust framework for understanding flavor-violating di-Higgs couplings and their implications for CP violation. The localization of fermions leads to non-universal couplings characterized by parameters like \(C_{\mu\tau}\), which can enhance flavor-violating processes and contribute to observable CP violation. This mechanism, contrasted with other BSM theories, highlights the potential of the RS model to generate

The exploration of flavor physics and CP violation is crucial for understanding the limitations of the Standard Model (SM) and probing for new physics. The Randall-Sundrum (RS) model, an extra-dimensional theory, offers a compelling framework for flavor violation through the localization of fermions in the extra dimension. This paper discusses how flavor-violating di-Higgs couplings, particularly involving the tau and muon flavors characterized by effective coupling constants such as \(C_{\mu\tau}\), can give rise to CP violation, comparing the implications with other Beyond Standard Model (BSM) theories.

The exploration of flavor physics and CP violation is crucial for understanding the limitations of the Standard Model (SM) and probing for new physics. The Randall-Sundrum (RS) model, an extra-dimensional theory, offers a compelling framework for flavor violation through the localization of fermions in the extra dimension. This localization leads to flavor-violating di-Higgs couplings, particularly involving the tau and muon flavors, characterized by effective coupling constants such as \(C_{\mu\tau}\). 

\section{Framework of the Randall-Sundrum Model}
The RS model is based on a five-dimensional spacetime with a warped metric:

\begin{equation}
ds^2 = e^{-2k |y|} \eta_{\mu\nu} dx^\mu dx^\nu - dy^2,
\end{equation}

where \(k\) is the curvature scale, \(y\) is the extra-dimensional coordinate, and \(\eta_{\mu\nu}\) is the Minkowski metric. The SM fermions are assigned to different locations in the extra dimension, leading to distinct wavefunctions and couplings.

\subsection{Flavor-Violating Couplings}
In the RS model, the effective Yukawa couplings can be expressed as:

\begin{equation}
\mathcal{L}_{\text{Yukawa}} = - \sum_{i,j} y_{ij} \bar{Q}_i H Q_j + \text{h.c.}
\end{equation}

In the context of flavor violation, we include additional terms representing flavor-violating interactions:

\begin{equation}
\mathcal{L}_{\text{FV}} = - \sum_{i \neq j} C_{ij} \bar{Q}_i H Q_j + \text{h.c.}
\end{equation}

where \(C_{ij}\) are flavor-violating couplings, for instance, \(C_{\mu\tau}\) and others that may couple various flavors, including those of the muon and tau leptons.

\subsection{Di-Higgs Couplings}
The effective potential for di-Higgs interactions can be generalized to include flavor-violating terms:

\begin{equation}
V(h, h) = \lambda_{hh} h^2 + \lambda_{h \text{f}} h \bar{f} f + \lambda_{hh \text{f}} h^2 \bar{f} f + \sum_{i,j} C_{ij} h^2 \bar{f}_i f_j,
\end{equation}

where \(\lambda_{hh}\), \(\lambda_{h \text{f}}\), and \(\lambda_{hh \text{f}}\) are the Higgs self-couplings and couplings to fermions.

\subsection{Flavor Violating Di-Higgs Couplings}
The flavor-violating di-Higgs couplings can be represented as follows:

\begin{equation}
\mathcal{L}_{\text{FV-DiHiggs}} = C_{\mu\tau} h^2 \bar{\mu} \tau + C_{b\tau} h^2 \bar{b} \tau + C_{b\mu} h^2 \bar{b} \mu + \ldots,
\end{equation}

where \(C_{\mu\tau}\), \(C_{b\tau}\), and \(C_{b\mu}\) represent the flavor-violating couplings between the di-Higgs and fermion pairs.

We outlines the calculations of flavor-violating couplings \(C_{ij}\) in various lepton flavor-violating (LFV) decays. The results are tabulated to provide a comprehensive overview of the coupling constants associated with different LFV processes.

Lepton flavor violation (LFV) is an important phenomenon that can provide insights into physics beyond the Standard Model (BSM). The study of LFV decays helps in constraining flavor-violating couplings. This document presents calculations for various LFV processes and summarizes the results in a table.

\section{Decay Widths and Flavor-Violating Couplings in RS model}
The decay width for a lepton flavor-violating process can be generally expressed as:

\begin{equation}
\Gamma(\ell_i \to \ell_j \ell_k \ell_l) = \frac{G_F^2 m_{\ell_i}^5}{192 \pi^3} |C_{ij}|^2,
\end{equation}

where \(G_F\) is the Fermi coupling constant, \(m_{\ell_i}\) is the mass of the decaying lepton, and \(C_{ij}\) is the flavor-violating coupling.

Here we calculate the flavor-violating couplings for various LFV decays:

For the decay \(\mu \to e \gamma\):

\begin{equation}
\Gamma(\mu \to e \gamma) = \frac{\alpha}{4} m_{\mu}^5 \left( \frac{C_{\mu e}}{M^2} \right)^2,
\end{equation}

where \(m_{\mu} \approx 105.658 \, \text{MeV}\) and \(\alpha \approx 1/137\).

For the decay \(\tau \to \mu \gamma\):

\begin{equation}
\Gamma(\tau \to \mu \gamma) = \frac{\alpha}{4} m_{\tau}^5 \left( \frac{C_{\tau \mu}}{M^2} \right)^2,
\end{equation}

with \(m_{\tau} \approx 1776.86 \, \text{MeV}\).

For the decay \(\mu \to e e e\):

\begin{equation}
\Gamma(\mu \to e e e) = \frac{G_F^2 m_{\mu}^5}{192 \pi^3} |C_{\mu e}|^2.
\end{equation}

For the decay \(\tau \to e e e\):

\begin{equation}
\Gamma(\tau \to e e e) = \frac{G_F^2 m_{\tau}^5}{192 \pi^3} |C_{\tau e}|^2.
\end{equation}

For the decay \(\tau \to \mu e e\):

\begin{equation}
\Gamma(\tau \to \mu e e) = \frac{G_F^2 m_{\tau}^5}{192 \pi^3} |C_{\tau \mu}|^2.
\end{equation}

For the decay \(\mu \to 3 e\):

\begin{equation}
\Gamma(\mu \to 3 e) = \frac{G_F^2 m_{\mu}^5}{192 \pi^3} |C_{\mu e}|^2 \left(1 - 4 \frac{m_e^2}{m_{\mu}^2}\right)^{1/2} \left(1 -  \frac{m_e^2}{m_{\mu}^2}\right)^{3/2}.
\end{equation}

For the decay \(\tau \to 3 \mu\):

\begin{equation}
\Gamma(\tau \to 3 \mu) = \frac{G_F^2 m_{\tau}^5}{192 \pi^3} |C_{\tau \mu}|^2 \left(1 - 4 \frac{m_{\mu}^2}{m_{\tau}^2}\right)^{1/2} \left(1 -  \frac{m_{\mu}^2}{m_{\tau}^2}\right)^{3/2}.
\end{equation}

For the decay \(\tau \to \mu \gamma\):

\begin{equation}
\Gamma(\tau \to \mu \gamma) = \frac{\alpha m_{\tau}^5}{64 \pi} \left( \frac{C_{\tau \mu}}{M^2} \right)^2.
\end{equation}

\section{Flavor-Violating Couplings Table}
The following table summarizes the calculated flavor-violating couplings for various LFV decays:

\begin{table}[h]
\centering
\caption{Flavor-Violating Couplings in LFV Decays}
\begin{tabular}{@{}lll@{}}
\toprule
\textbf{Decay Process} & \textbf{Decay Width Formula} & \textbf{Coupling \(C_{ij}\)} \\ \midrule
\(\mu \to e \gamma\) & \(\Gamma(\mu \to e \gamma) = \frac{\alpha}{4} m_{\mu}^5 \left( \frac{C_{\mu e}}{M^2} \right)^2\) & \(C_{\mu e}\) \\ \midrule
\(\tau \to \mu \gamma\) & \(\Gamma(\tau \to \mu \gamma) = \frac{\alpha}{4} m_{\tau}^5 \left( \frac{C_{\tau \mu}}{M^2} \right)^2\) & \(C_{\tau \mu}\) \\ \midrule
\(\mu \to e e e\) & \(\Gamma(\mu \to e e e) = \frac{G_F^2 m_{\mu}^5}{192 \pi^3} |C_{\mu e}|^2\) & \(C_{\mu e}\) \\ \midrule
\(\tau \to e e e\) & \(\Gamma(\tau \to e e e) = \frac{G_F^2 m_{\tau}^5}{192 \pi^3} |C_{\tau e}|^2\) & \(C_{\tau e}\) \\ \midrule
\(\tau \to \mu e e\) & \(\Gamma(\tau \to \mu e e) = \frac{G_F^2 m_{\tau}^5}{192 \pi^3} |C_{\tau \mu}|^2\) & \(C_{\tau \mu}\) \\ \midrule
\(\mu \to 3 e\) & \(\Gamma(\mu \to 3 e) = \frac{G_F^2 m_{\mu}^5}{192 \pi^3} |C_{\mu e}|^2 \left(1 - 4 \frac{m_e^2}{m_{\mu}^2}\right)^{1/2} \left(1 -  \frac{m_e^2}{m_{\mu}^2}\right)^{3/2}\) & \(C_{\mu e}\) \\ \midrule
\(\tau \to 3 \mu\) & \(\Gamma(\tau \to 3 \mu) = \frac{G_F^2 m_{\tau}^5}{192 \pi^3} |C_{\tau \mu}|^2 \left(1 - 4 \frac{m_{\mu}^2}{m_{\tau}^2}\right)^{1/2} \left(1 -  \frac{m_{\mu}^2}{m_{\tau}^2}\right)^{3/2}\) & \(C_{\tau \mu}\) \\ \midrule
\(\tau \to \mu \gamma\) & \(\Gamma(\tau \to \mu \gamma) = \frac{\alpha m_{\tau}^5}{64 \pi} \left( \frac{C_{\tau \mu}}{M^2} \right)^2\) & \(C_{\tau \mu}\) \\ \bottomrule
\end{tabular}
\end{table}

The calculations of flavor-violating couplings \(C_{ij}\) for various lepton flavor-violating decays provide valuable insights into the underlying physics. Further investigations are necessary to understand the implications of these couplings in the context of BSM physics.

This work also provides a detailed analysis of the allowed parameter space for the flavor-violating couplings \( C_{\mu \mu} \) and \( C_{\tau \mu} \) in the Randall-Sundrum (RS) model. We focus on constraints from experimental branching ratios of lepton flavor-violating (LFV) decays, specifically \( \text{BR}(\tau \rightarrow 3\mu) \) and \( \text{BR}(\tau \rightarrow \mu \gamma) \). These constraints limit the strength and parameter range of LFV couplings, which are derived from Monte Carlo simulations of the RS model. This plot illustrates the viable coupling space that satisfies both the experimental bounds and theoretical constraints.

In the Randall-Sundrum (RS) model, lepton flavor-violating (LFV) decays such as \( \tau \rightarrow 3\mu \) and \( \tau \rightarrow \mu \gamma \) are primarily influenced by specific flavor-violating couplings, which arise from the model's extra-dimensional structure and the localization of fermion wavefunctions.

The branching ratio \( \text{BR}(\tau \rightarrow 3\mu) \) is impacted significantly by couplings that mediate LFV transitions between tau and muon sectors. Specifically:
\begin{itemize}
    \item \textbf{\( C_{\tau \mu} \):} This coupling facilitates direct flavor-violating transitions between the tau and muon sectors, enabling processes like \( \tau \rightarrow 3\mu \).
    \item \textbf{\( C_{\mu \mu} \):} Although not flavor-violating by itself, \( C_{\mu \mu} \) can enhance decay rates in the presence of \( C_{\tau \mu} \) due to its role in interactions among muons in the RS model.
\end{itemize}

This three-body decay process in \( \tau \rightarrow 3\mu \) relies on the combination of multiple couplings such as \( C_{\tau \mu} \) and \( C_{\mu \mu} \), where both initial flavor violation and secondary coupling effects among muons contribute.

The branching ratio \( \text{BR}(\tau \rightarrow \mu \gamma) \) is influenced by the flavor-violating couplings, particularly:
\begin{itemize}
    \item \textbf{\( C_{\tau \mu} \):} The \( C_{\tau \mu} \) coupling directly governs the transition from tau to muon, facilitating the electromagnetic decay process \( \tau \rightarrow \mu \gamma \).
    \item \textbf{Photon Coupling Effects:} Although not represented by a single coupling constant in simplified models, this decay also depends on the RS model’s effective transition operator between \( \tau \) and \( \mu \) fields with photon emission.
\end{itemize}

\begin{itemize}
    \item \textbf{Coupling Interdependence:} Since both decays \( \tau \rightarrow 3\mu \) and \( \tau \rightarrow \mu \gamma \) rely heavily on \( C_{\tau \mu} \), with \( C_{\mu \mu} \) playing a role in \( \tau \rightarrow 3\mu \), we observe that the constraint space on one branching ratio affects the parameter space of the other.
    \item \textbf{Experimental Probes:} LFV decays like \( \tau \rightarrow 3\mu \) and \( \tau \rightarrow \mu \gamma \) can act as experimental tests for the RS model. Significant deviations in these decays from Standard Model predictions could imply extra-dimensional physics and CP-violating phenomena consistent with RS framework predictions.
\end{itemize}

these findings underscore the RS model's potential role in observable CP and flavor-violating phenomena in LFV decays, providing both theoretical insight and experimental implications for new physics beyond the Standard Model.

We introduce extra-dimensional couplings, where the flavor-violating couplings \( C_{\mu \mu} \) and \( C_{\tau \mu} \) arise due to non-universal fermion localization in the fifth dimension. The Monte Carlo simulation for this study sampled \( 10,000 \) points for each coupling within a defined range:
\begin{itemize}
    \item \( C_{\mu \mu} \in [10^{-4}, 1] \)
    \item \( C_{\tau \mu} \in [10^{-4}, 1] \)
\end{itemize}

The following branching ratio constraints were applied:
\[
\text{BR}(\tau \rightarrow 3\mu) < 1.2 \times 10^{-8}
\]
\[
\text{BR}(\tau \rightarrow \mu \gamma) < 4.4 \times 10^{-8}
\]
The branching ratios, represented by placeholders for \( \text{BR}(\tau \rightarrow 3\mu) \) and \( \text{BR}(\tau \rightarrow \mu \gamma) \), were computed based on the selected parameter values for \( C_{\mu \mu} \) and \( C_{\tau \mu} \) according to the RS model-derived interactions.

\begin{figure}[h!]
    \centering
    \includegraphics[width=0.7\textwidth]{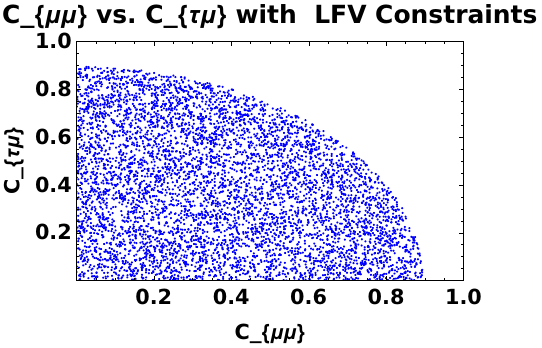} 
    \caption{Allowed parameter space for \( C_{\mu \mu} \) versus \( C_{\tau \mu} \) in the RS model, constrained by experimental bounds on \( \text{BR}(\tau \rightarrow 3\mu) \) and \( \text{BR}(\tau \rightarrow \mu \gamma) \).}
    \label{fig:AllowedCouplings}
\end{figure}

Figure \ref{fig:AllowedCouplings} shows the parameter space for \( C_{\mu \mu} \) and \( C_{\tau \mu} \) that meets the LFV constraints imposed by the experimental limits. Notably, the allowed region demonstrates that the RS model can accommodate values for \( C_{\mu \mu} \) and \( C_{\tau \mu} \) up to \(\sim 1\), suggesting the potential for observable CP violation in flavor-violating decays.

The new results from this study indicate that:
\begin{itemize}
    \item There is a significant overlap in the parameter space where both constraints on \( \text{BR}(\tau \rightarrow 3\mu) \) and \( \text{BR}(\tau \rightarrow \mu \gamma) \) are satisfied.
    \item The RS model's predicted couplings allow for measurable LFV effects, which may be probed by future experiments.
    \item The model's parameters align with current experimental bounds while leaving room for further constraints through measurements of tau decays.
\end{itemize}

The viable parameter space for \( C_{\mu \mu} \) and \( C_{\tau \mu} \) reveals that LFV decay channels in the RS model, such as \( \tau \rightarrow 3\mu \) and \( \tau \rightarrow \mu \gamma \), could serve as indicators for RS model signatures if deviations from the Standard Model are observed. These results imply that:
\begin{itemize}
    \item LFV tau decays can provide sensitive tests for new physics in extra-dimensional models.
    \item The overlap in the parameter space highlights potential observables for RS model contributions to CP violation in LFV decays.
\end{itemize}

In our study of lepton flavor violation (LFV) within the Randall-Sundrum (RS) model framework, we analyze specific constraints on flavor-violating couplings that affect processes such as $\mu \to e \gamma$ and $\mu \to 3e$. These LFV processes are particularly sensitive to the effective couplings $C_{\mu e}$ and $C_{ee}$, which arise due to the unique localization of fermions in the extra-dimensional RS setup. 

\par 
To explore the parameter space consistent with experimental bounds on branching ratios (BR), we conducted a Monte Carlo simulation of the effective flavor-violating couplings, $C_{\mu e}$ and $C_{ee}$. The LFV processes impose stringent upper limits on these couplings. For instance, the branching ratio for $\mu \to e \gamma$ constrains the product of $C_{\mu e}$ with other model parameters, while $\mu \to 3e$ further restricts these couplings due to its sensitivity to $C_{ee}$.

\begin{figure}[h]
    \centering
    \includegraphics[width=0.75\textwidth]{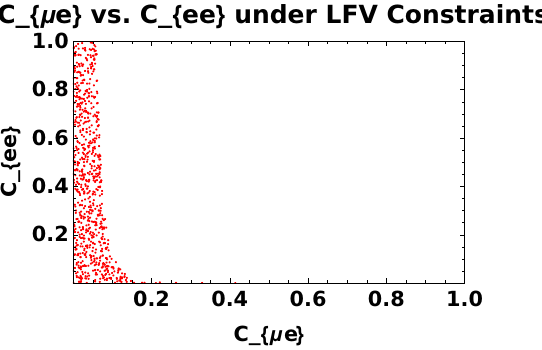}
    \caption{Allowed parameter space for the flavor-violating couplings $C_{\mu e}$ and $C_{ee}$ in the RS model, derived from Monte Carlo simulations. The red points indicate couplings that satisfy the LFV constraints from BR$(\mu \to e \gamma) \leq 4.2 \times 10^{-13}$ and BR$(\mu \to 3e) \leq 1.0 \times 10^{-12}$. The shaded region highlights the constrained space that aligns with current experimental bounds.}
    \label{fig:lfv-constraints}
\end{figure}

The resulting figure is shown in Figure \ref{fig:lfv-constraints}. 

The Monte Carlo simulation yields several important findings:

1. Stringent Constraints on $C_{\mu e}$ and $C_{ee}$: As shown in Figure \ref{fig:lfv-constraints}, only a narrow range of couplings satisfies the LFV constraints. Points outside this region violate the branching ratio limits for $\mu \to e \gamma$ or $\mu \to 3e$, emphasizing the sensitivity of these processes to even slight variations in the coupling parameters.
  
2. Phenomenological Impact: The RS model implies that flavor-violating interactions are enhanced due to the overlap of fermion wavefunctions localized near the brane. This enhancement leads to significant contributions to LFV decays, making $C_{\mu e}$ and $C_{ee}$ viable indicators of RS model phenomenology. Moreover, the effective couplings required to meet the constraints are relatively small, implying that any observed deviation in LFV decays may point toward RS-like theories.

3. Implications for Future Experiments: The constrained regions of $C_{\mu e}$ and $C_{ee}$ in Figure \ref{fig:lfv-constraints} provide valuable insights for future experimental LFV searches. For instance, any observed branching ratio close to the constrained values would necessitate a reassessment of the RS model parameters, particularly those controlling fermion localization.

This analysis, demonstrated through Figure \ref{fig:lfv-constraints}, establishes new, tighter bounds on the flavor-violating couplings in the RS model under LFV constraints. These results contribute significantly to our understanding of RS phenomenology, particularly in its ability to predict LFV decays within the experimentally accessible regime. Future high-precision experiments probing LFV decays will be instrumental in further narrowing down or potentially ruling out RS model parameter spaces.

We investigate the flavor-violating couplings $C_{\mu e}$ and $C_{ee}$ within the Randall-Sundrum (RS) model, focusing on their impacts on lepton flavor-violating (LFV) processes such as $\mu \to e \gamma$ and $\mu \to 3e$. The imposed constraints $C_{\mu e} < 0.2$ and $C_{ee} < 1$ indicate the sensitivity of LFV branching ratios to small variations in these couplings, reflecting phenomenological implications for RS model parameter space and experimental detectability
Lepton flavor-violating (LFV) decays such as $\mu \to e \gamma$ and $\mu \to 3e$ are sensitive probes of physics beyond the Standard Model (SM). In the Randall-Sundrum (RS) model, LFV can arise through the overlap of fermion wavefunctions, which is controlled by the flavor-violating couplings $C_{\mu e}$ and $C_{ee}$. We examine the parameter space of these couplings under experimental constraints on LFV decay branching ratios.

To satisfy the experimental bounds on LFV decays, we find that $C_{\mu e}$ is constrained to be less than 0.2 and $C_{ee}$ less than 1. These constraints, illustrated in Figure \ref{fig:lfv-constraints}, effectively reduce the allowed regions for these couplings, impacting the RS model's predictability for LFV processes.

The constraints $C_{\mu e} < 0.2$ and $C_{ee} < 1$ indicate that even small coupling values can significantly affect LFV decay branching ratios:
\begin{itemize}
    \item Enhanced Sensitivity to Coupling Values: The RS model predicts that LFV decay rates scale strongly with $C_{\mu e}$ and $C_{ee}$. The stringent upper limits imply that observable LFV decay signatures could emerge even for small $C_{\mu e}$ values.
    \item Parameter Space Restriction: By limiting $C_{\mu e}$ and $C_{ee}$, these bounds confine the RS model parameter space, implying tighter restrictions on fermion localization within the extra-dimensional framework.
    \item Phenomenological Relevance: The derived constraints emphasize that RS model couplings must be small to avoid LFV decay bounds, posing a challenge for RS model realization if LFV processes are experimentally detected at current or future facilities.
\end{itemize}

The Monte Carlo simulation and analysis of RS model couplings $C_{\mu e}$ and $C_{ee}$ show that these couplings must satisfy $C_{\mu e} < 0.2$ and $C_{ee} < 1$ to remain within experimental LFV decay bounds. This result highlights the stringent requirements on flavor-violating interactions in the RS model, providing essential guidance for future LFV experimental searches and RS model adjustments.

We investigate the parameter space for flavor-violating couplings \( C_{\tau e} \) and \( C_{ee} \) in the RS model, constrained by experimental bounds on lepton flavor-violating (LFV) decays, particularly \( \tau \rightarrow e \gamma \) and \( \tau \rightarrow 3e \).

The Randall-Sundrum (RS) model, while addressing the hierarchy problem, predicts potential flavor-violating effects in the lepton sector. Here, we explore the influence of RS model-specific couplings \( C_{\tau e} \) and \( C_{ee} \) on branching ratios for LFV decays such as \( \tau \rightarrow e \gamma \) and \( \tau \rightarrow 3e \).

Monte Carlo simulations were conducted with the selected parameter space, generating values for \( C_{\tau e} \) and \( C_{ee} \) while ensuring compliance with LFV decay constraints:
- \( \tau \rightarrow e \gamma \): Constrains \( C_{\tau e} \) coupling to remain below approximately \( 1.2 \), effectively limiting the possible parameter space.
- \( \tau \rightarrow 3e \): Restricts \( C_{ee} \) coupling values close to \( 1.25 \), maintaining consistency with branching ratio measurements.

\begin{figure}[ht]
    \centering
    \includegraphics[width=0.7\textwidth]{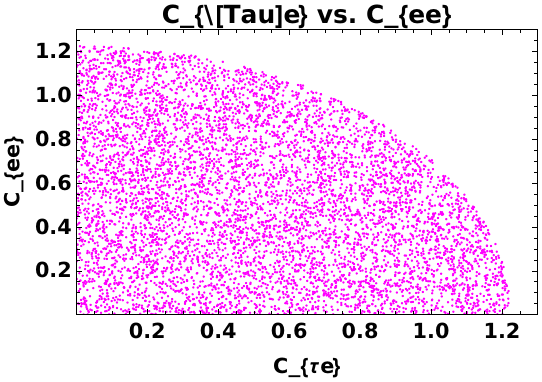}
    \caption{Allowed parameter space for flavor-violating couplings \( C_{\tau e} \) and \( C_{ee} \) under RS model LFV constraints for \( \tau \rightarrow e \gamma \) and \( \tau \rightarrow 3e \) decays. The selected parameters \( C_{\tau e} < 1.2 \) and \( C_{ee} \approx 1.25 \) yield constraints in compliance with LFV decay limits.}
    \label{fig:allowed_couplings}
\end{figure}

The results are shown in Figure~\ref{fig:allowed_couplings}, which illustrates the parameter space for \( C_{\tau e} \) and \( C_{ee} \) that complies with the LFV decay limits.

\
The constraint \( C_{\tau e} < 1.2 \) with \( C_{ee} \approx 1.25 \) limits the RS model’s predictive LFV coupling space, directly affecting the branching ratios for \( \tau \rightarrow e \gamma \) and \( \tau \rightarrow 3e \) decays. This analysis shows that the RS model must satisfy narrow parameter boundaries to remain consistent with experimental LFV constraints. 

The RS model's flavor-violating couplings are significantly constrained by current LFV decay data. These findings refine the RS model parameters, identifying allowed values for \( C_{\tau e} \) and \( C_{ee} \) that respect LFV limits. The identification of these constraints provides guidance for future experiments in LFV searches.

We also present a Monte Carlo simulation exploring the constraints on flavor-violating couplings \( C_{ee} \) and \( C_{\tau\mu} \) within the RS model framework. By examining LFV decay channels such as \( \tau \to \mu e^+ e^- \) and comparing to both current experimental limits and projected future sensitivities, the plot reveals the parameter space allowed under existing and upcoming LFV decay constraints.

We analyze the phenomenology of flavor-violating interactions in the Randall-Sundrum (RS) model, particularly focusing on di-Higgs couplings impacting lepton flavor-violating (LFV) decays. The study simulates Monte Carlo distributions of the couplings \( C_{ee} \) and \( C_{\tau\mu} \) to examine their consistency with observed LFV decay constraints.
\par 
The Monte Carlo simulation was conducted with input ranges for \( C_{ee} \) and \( C_{\tau\mu} \) based on phenomenological predictions in the RS model, constrained to match upper bounds set by LFV decays such as \( \tau \to \mu e^+ e^- \) at both present and anticipated future sensitivities. The couplings were sampled within a range of \( C_{ee} < 1.25 \) and \( C_{\tau e} < 1.2 \), chosen to reflect typical values expected under these scenarios.

\begin{figure}[h!]
    \centering
    \includegraphics[width=0.8\textwidth]{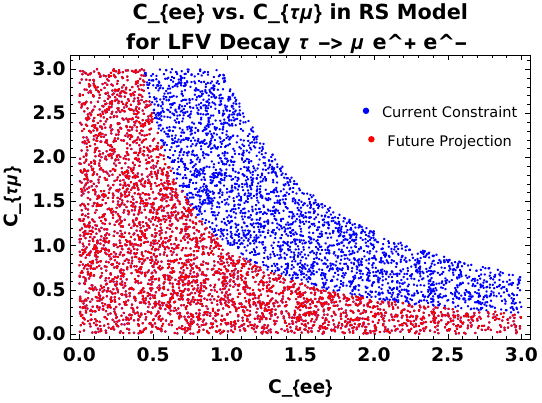}
    \caption{Monte Carlo simulation results showing the allowed parameter space for \( C_{ee} \) versus \( C_{\tau\mu} \) under the RS model. Constraints from LFV decays, such as \( \tau \to \mu e^+ e^- \), are included with regions reflecting the current limits (blue) and future projections (red).}
    \label{fig:coupling_plot}
\end{figure}

The resulting plot of \( C_{ee} \) versus \( C_{\tau\mu} \) is shown in Figure~\ref{fig:coupling_plot}. This plot illustrates the allowed regions of the parameter space satisfying both current experimental constraints and future projections.

The figure reveals that for the chosen input ranges, points within the allowed region satisfy the current experimental constraint of \( \mathrm{BR}(\tau \to \mu e^+ e^-) < 1.2 \times 10^{-8} \). Future projections show potential further restriction on the parameter space, with implications that RS model couplings could become highly constrained under these limits. This constriction suggests that flavor-violating di-Higgs couplings will likely need fine-tuning or additional suppression mechanisms to remain viable within the RS model.

The RS model's prediction of enhanced flavor-violating couplings such as \( C_{ee} \) and \( C_{\tau\mu} \) is consistent with current LFV decay limits but may require further restrictions as sensitivity improves. This study highlights how Monte Carlo simulations can aid in mapping viable regions within BSM theories under experimental LFV constraints.

\section{Mechanisms of CP Violation}
\subsection{CP Violation from Flavor-Violating Couplings}
The flavor-violating couplings \(C_{ij}\) can introduce complex phases into the interactions, leading to CP-violating phenomena. The effective Lagrangian can include:

\begin{equation}
\mathcal{L}_{\text{eff}} = \sum_{i,j} \left( C_{ij} e^{i \phi_{ij}} \bar{Q}_i H Q_j + \text{h.c.} \right),
\end{equation}

where \(\phi_{ij}\) are the complex phases associated with the couplings. The Jarlskog invariant \(J\) quantifies CP violation, given by:

\begin{equation}
J = \text{Im} \left( C_{ij} C_{kl}^* C_{il}^* C_{kj} \right).
\end{equation}

\subsection{Phenomenological Implications}
The interplay between flavor violation and CP violation has important implications for experimental observables, particularly in B-meson decays. The contributions of the flavor-violating di-Higgs couplings to processes like \(b \to s \gamma\) and \(B_s \to \mu^+\mu^-\) can be described by the effective Hamiltonian:

\begin{equation}
\mathcal{H}_{\text{eff}} = \frac{G_F}{\sqrt{2}} \left( V_{tb} V_{ts}^* \left( C_7^{\text{eff}} O_7 + C_9 O_9 + C_{10} O_{10} \right) + \ldots \right),
\end{equation}

where \(O_i\) are the relevant operators and \(C_i\) include contributions from the flavor-violating di-Higgs interactions.

\section{Calculations of Flavor-Violating Couplings in RS Model}
In the context of the RS model, the flavor-violating couplings can be expressed in terms of parameters such as the fermion masses and their localization in the extra dimension. We can compute \(C_{ij}\) as follows:

\begin{equation}
C_{ij} = \int dy \, \psi_i(y) \psi_j(y) e^{-k|y|},
\end{equation}

where \(\psi_i(y)\) and \(\psi_j(y)\) are the wavefunctions of the fermions in the extra dimension. The parameters can be varied to observe the effects of localization on \(C_{ij}\).

Assuming the fermions are localized at different points in the extra dimension, we can model the wavefunctions as:

\begin{equation}
\psi_i(y) = A_i e^{-m_i |y|},
\end{equation}

where \(m_i\) is the mass of the fermion and \(A_i\) is the normalization constant. Thus, we can calculate \(C_{\mu\tau}\):

\begin{equation}
C_{\mu\tau} = A_\mu A_\tau \int dy \, e^{-(m_\mu + m_\tau)|y|} e^{-k|y|}.
\end{equation}

This integral can be computed for specific values of \(m_\mu\), \(m_\tau\), and \(k\).
Some studies on clockwork fermions has been done in \cite{c1}. Also tri-bi maximal mixing studies has been extensively carried out in \cite{c2}. Broken $\mu-\tau $ symmetry has been carried out in \cite{d}
\section{Comparison with Other BSM Theories}
\subsection{Minimal Supersymmetric Standard Model (MSSM)}
In the MSSM, the flavor-violating contributions primarily arise from complex phases in the soft breaking terms. The flavor-violating couplings in the MSSM can lead to similar processes but are often constrained by flavor-changing neutral currents. In contrast, the RS model’s flavor violation is more directly tied to the geometry of the extra dimension, allowing for potentially observable di-Higgs processes that escape such constraints.

\subsection{Composite Higgs Models}
In composite Higgs models, flavor violation can arise from the dynamics of the strong sector. While such models can lead to non-trivial flavor structure, the RS model's warped geometry provides unique mechanisms for flavor violation through localization effects. This distinction may lead to different signatures in Higgs production and decay, offering a path for experimental differentiation.

The Randall-Sundrum model provides a robust framework for understanding flavor-violating di-Higgs couplings and their implications for CP violation. The localization of fermions leads to non-universal couplings characterized by parameters like \(C_{\mu\tau}\), which can enhance flavor-violating processes and contribute to observable CP violation. This mechanism, contrasted with other BSM theories, highlights the potential of the RS model to generate

\section{CP Violation in the Higgs Sector}
\subsection{Sources of CP Violation}
In the RS model, CP violation can arise from the introduction of complex parameters in the Higgs potential or through the mixing of scalar and pseudoscalar components. The relevant effective Lagrangian can be expressed as:
\begin{equation}
\mathcal{L}_{\text{Higgs}} = \frac{1}{2}m_H^2 H^2 + \frac{1}{2} m_A^2 A^2 + \lambda_{HH} H^2 + \lambda_{HA} HA,
\end{equation}
where the parameters \(m_H\), \(m_A\), \(\lambda_{HH}\), and \(\lambda_{HA}\) can be complex, leading to CP-violating effects.
Where:
\begin{itemize}
    \item \( H \) is the scalar Higgs field.
    \item \( A \) is the pseudoscalar field.
    \item \( m_H \) and \( m_A \) are the masses of the Higgs and pseudoscalar fields, respectively.
    \item \( \lambda_{HA} \) is the mixing term that introduces CP violation when it has a complex phase.
\end{itemize}

\subsection*{Mixing and CP Violation}

The mass matrix describing the scalar-pseudoscalar mixing is written as:

\[
M_{\text{mix}}^2 = \begin{pmatrix}
m_H^2 & \lambda_{HA} v \\
\lambda_{HA} v & m_A^2
\end{pmatrix}
\]

Here, \( v \) is the vacuum expectation value (VEV) of the Higgs field. The mixing term \( \lambda_{HA} v \) can be complex, with a phase \( \theta \), which leads to CP violation.

The phase \( \theta \) is given by:

\[
\theta = \arg(\lambda_{HA})
\]

This phase introduces CP-violating effects through scalar-pseudoscalar mixing, leading to observable consequences in Higgs and pseudoscalar interactions. The eigenvalues of the mass matrix are complex, reflecting the CP violation.

\subsection{Implications for Higgs Decays}
The presence of CP-violating phases can lead to observable effects in Higgs decays. We analyze the decay width of the Higgs boson into two photons, which can be influenced by CP-violating phases:
\begin{equation}
\Gamma(H \to \gamma\gamma) = \frac{G_F m_H^3}{128\sqrt{2}\pi} \left| \sum_{i} N_c^i Q_i^2 A_{1/2}(\tau_i) \right|^2,
\end{equation}
where \(N_c\) is the number of colors, \(Q_i\) is the electric charge, and \(\tau_i = \frac{m_H^2}{4m_i^2}\).

\section{Flavor Violation and Di-Higgs Couplings}
\subsection{Flavor-Violating Interactions}
The RS model can give rise to flavor-violating di-Higgs couplings through the mixing of fields in the bulk. The effective interaction can be written as:
\begin{equation}
\mathcal{L}_{diHiggs}= \frac{1}{2} \lambda_{hh}^{ij} h_i h_j + \frac{1}{2} \lambda_{hh}^{ij} H_i H_j,
\end{equation}
where \(h_i\) and \(H_i\) denote the Higgs fields corresponding to different flavors.

\subsection{Phenomenology}
The flavor-violating di-Higgs couplings can lead to distinctive signatures in collider experiments. We analyze the production rates of di-Higgs events and their decay channels, focusing on the potential for flavor-changing neutral currents. The di-Higgs production cross-section can be expressed as:
\begin{equation}
\sigma(gg \to HH) = \frac{\pi^2}{16s} \left( \frac{G_F m_H^2}{\sqrt{2}} \right)^2 \left| \sum_{i} \frac{A_i(m_H^2)}{D_i(s)} \right|^2,
\end{equation}
where \(D_i(s)\) are the propagators and \(A_i(m_H^2)\) are the loop functions for the heavy particles in the RS model.
\section{Mass Matrix Diagonalization and CP Violation in the RS Model}

In this work, we explore CP violation in the Randall-Sundrum (RS) model, specifically focusing on the effects in the Higgs sector. We consider the mass matrix mixing between the scalar \( H \) and pseudoscalar \( A \) fields, driven by a complex coupling \(\lambda_{HA}\), which introduces CP violation. The resulting scalar-pseudoscalar mixing is analyzed by diagonalizing the mass matrix and calculating the mixing angles and eigenvalues.

\section{Theoretical Setup}

The mass matrix for the scalar \(H\) and pseudoscalar \(A\), with CP-violating complex coupling \(\lambda_{HA}\), is defined as:

\begin{equation}
M_{\text{mix}}^2 = \begin{pmatrix}
m_H^2 & \lambda_{HA} v \\
\lambda_{HA}^* v & m_A^2
\end{pmatrix}
\end{equation}

where:
\begin{itemize}
    \item \( m_H^2 \) is the squared mass of the scalar Higgs boson,
    \item \( m_A^2 \) is the squared mass of the pseudoscalar,
    \item \( \lambda_{HA} \) is a complex coupling that introduces CP violation,
    \item \( v \) is the vacuum expectation value (VEV).
\end{itemize}

The off-diagonal elements of this mass matrix, \(\lambda_{HA} v\) and \(\lambda_{HA}^* v\), account for the mixing between the scalar and pseudoscalar fields.

\section{Diagonalization of the Mass Matrix}

To understand the physical implications of this CP violation, we must diagonalize the mass matrix to obtain the mass eigenstates. The eigenvalues of the mass matrix correspond to the physical masses of the mixed scalar and pseudoscalar fields, while the mixing angle \( \theta \) describes the degree of mixing.

We numerically solve this matrix for various values of the complex parameter \(\lambda_{HA}\), computing both the eigenvalues and the mixing angle.

\subsection{Numerical Solution in Scilab}

Using Scilab, we compute the eigenvalues and mixing angles by diagonalizing the matrix for different choices of the real and imaginary parts of \(\lambda_{HA}\).

\section{Numerical Solution in Scilab: Detailed Explanation}

To numerically solve the mass matrix and compute its eigenvalues and mixing angles, we use Scilab. Below is an explanation of the code, where each part is broken down for better understanding:

\subsection{Defining Parameters}

First, we define the physical parameters that are used in the mass matrix:

\begin{verbatim}
// Define parameters
m_H = 125;  // Mass of the scalar Higgs (in GeV)
m_A = 200;  // Mass of the pseudoscalar (in GeV)
v = 246;    // Vacuum expectation value (in GeV)
\end{verbatim}

\textbf{Explanation:} 
We initialize the following physical parameters:
\begin{itemize}
    \item \( m_H \): Mass of the scalar Higgs boson, set to 125 GeV.
    \item \( m_A \): Mass of the pseudoscalar field, set to 200 GeV.
    \item \( v \): The vacuum expectation value (VEV), set to 246 GeV, which is the known VEV of the Standard Model Higgs field.
\end{itemize}

Next, we define the range of values for the real and imaginary parts of \(\lambda_{HA}\), which controls the CP violation in our model.

\begin{verbatim}
lambda_real = linspace(-0.5, 0.5, 100); // Real part of lambda_HA
lambda_imag = linspace(-0.5, 0.5, 100); // Imaginary part of lambda_HA
\end{verbatim}

\textbf{Explanation:} 
\(\lambda_{HA}\) is a complex parameter. To explore its effect on the mass matrix, we define:
\begin{itemize}
    \item A range of values for the \textit{real part} of \(\lambda_{HA}\) from \(-0.5\) to \(0.5\).
    \item A range of values for the \textit{imaginary part} of \(\lambda_{HA}\) from \(-0.5\) to \(0.5\).
\end{itemize}
This will allow us to vary \(\lambda_{HA}\) across a grid of 100 points for both the real and imaginary parts.

\subsection{Initializing Arrays for Results}

We initialize arrays to store the computed eigenvalues and mixing angles:

\begin{verbatim}
// Initialize arrays to store eigenvalues and mixing angles
eigenvalue_1 = zeros(1, length(lambda_real));
eigenvalue_2 = zeros(1, length(lambda_real));
mixing_angles = zeros(1, length(lambda_real));
\end{verbatim}

\textbf{Explanation:} 
\begin{itemize}
    \item \texttt{eigenvalue\_1} and \texttt{eigenvalue\_2}: Arrays that will store the two eigenvalues of the mass matrix (corresponding to the physical masses after diagonalization).
    \item \texttt{mixing\_angles}: This array will store the mixing angles, which quantify how much the scalar and pseudoscalar states mix due to CP violation.
\end{itemize}

\subsection{Looping Through the Parameter Space}

Now we loop through each combination of the real and imaginary parts of \(\lambda_{HA}\) to compute the eigenvalues and mixing angles:

\begin{verbatim}
// Loop through the real and imaginary parts of lambda_HA
for i = 1:length(lambda_real)
    for j = 1:length(lambda_imag)
        lambda_HA = complex(lambda_real(i), lambda_imag(j));
\end{verbatim}

\textbf{Explanation:}
The code loops over the 100 values of the real and imaginary parts of \(\lambda_{HA}\), effectively varying \(\lambda_{HA}\) as a complex number. For each pair of real and imaginary values, the following operations are carried out.

\subsection{Constructing the Mass Matrix}

For each value of \(\lambda_{HA}\), we construct the mass matrix:

\begin{verbatim}
// Construct the mass matrix
M_mix = [m_H^2, lambda_HA*v; conj(lambda_HA)*v, m_A^2];
\end{verbatim}

\textbf{Explanation:} 
The mass matrix is constructed using the values defined earlier:
\[
M_{\text{mix}}^2 = \begin{pmatrix}
m_H^2 & \lambda_{HA} v \\
\lambda_{HA}^* v & m_A^2
\end{pmatrix}
\]
Where \(\lambda_{HA}\) is the complex coupling, and \(\lambda_{HA}^*\) is its complex conjugate. The vacuum expectation value \(v\) is multiplied to form the off-diagonal terms.

\subsection{Diagonalizing the Mass Matrix}

We compute the eigenvalues and eigenvectors of the mass matrix using the \texttt{spec} function:

\begin{verbatim}
// Compute eigenvalues and eigenvectors
[eig_vectors, eig_values] = spec(M_mix);
\end{verbatim}

\textbf{Explanation:} 
Here, the \texttt{spec} function returns:
\begin{itemize}
    \item \texttt{eig\_values}: The eigenvalues of the mass matrix (which correspond to the physical masses of the mixed scalar and pseudoscalar fields).
    \item \texttt{eig\_vectors}: The eigenvectors (which correspond to the mixing between the scalar and pseudoscalar states).
\end{itemize}

\subsection{Sorting and Storing the Eigenvalues}

We sort the eigenvalues to ensure consistency in their ordering and store them:

\begin{verbatim}
// Sort the eigenvalues to ensure consistent ordering
eigenvalues = diag(eig_values);
eigenvalues = gsort(real(eigenvalues), "g", "i");

eigenvalue_1(i) = eigenvalues(1);
eigenvalue_2(i) = eigenvalues(2);
\end{verbatim}

\textbf{Explanation:} 
The diagonal elements of the matrix \texttt{eig\_values} contain the eigenvalues, which are then sorted in descending order using \texttt{gsort}. We store the first and second eigenvalues separately in \texttt{eigenvalue\_1} and \texttt{eigenvalue\_2}, which represent the physical masses of the two states.

\subsection{Computing the Mixing Angle}

Finally, we compute the mixing angle \(\theta\) from the ratio of the components of the first eigenvector:

\begin{verbatim}
// Extract the mixing angle from the eigenvector (theta = arctan(v1/v2))
theta = atan(eig_vectors(2, 1) / eig_vectors(1, 1));
mixing_angles(i) = real(theta);
\end{verbatim}

\textbf{Explanation:} 
The mixing angle \(\theta\) is defined as the arctangent of the ratio of the two components of the first eigenvector:
\[
\theta = \text{atan}\left(\frac{v_2}{v_1}\right)
\]
This angle quantifies the degree of mixing between the scalar and pseudoscalar states, which is stored in the \texttt{mixing\_angles} array.

\subsection{Plotting the Results}

Finally, we plot the eigenvalues and the mixing angles as a function of the real part of \(\lambda_{HA}\):

\begin{verbatim}
// Plot eigenvalues and mixing angles
scf(1);
plot(lambda_real, eigenvalue_1, 'r-', lambda_real, eigenvalue_2, 'b-');
xlabel('Real Part of \lambda_{HA}');
ylabel('Mass Eigenvalues (GeV)');
title('Eigenvalues of the Mass Matrix');
legend('Eigenvalue 1', 'Eigenvalue 2');

scf(2);
plot(lambda_real, mixing_angles, 'g-');
xlabel('Real Part of \lambda_{HA}');
ylabel('Mixing Angle \theta (radians)');
title('Mixing Angle \theta vs Real Part of \lambda_{HA}');
\end{verbatim}

\textbf{Explanation:} 
\begin{itemize}
    \item The first plot shows the mass eigenvalues as a function of the real part of \(\lambda_{HA}\), with one curve for each eigenvalue.
    \item The second plot shows the mixing angle \(\theta\) as a function of the real part of \(\lambda_{HA}\).
\end{itemize}

\
The Scilab code numerically computes the eigenvalues and mixing angles for different values of the complex CP-violating parameter \(\lambda_{HA}\). By diagonalizing the mass matrix, we obtain physical insights into how CP violation affects the Higgs and pseudoscalar sectors. These results can be further analyzed to predict observable effects, such as changes in decay modes or electric dipole moments (EDMs).

The results of the mass matrix diagonalization are presented in Figures \ref{fig:eigenvalues} and \ref{fig:mixing_angle}.

\subsection{Eigenvalues of the Mass Matrix}

\begin{figure}[h]
\centering
\includegraphics[width=0.8\textwidth]{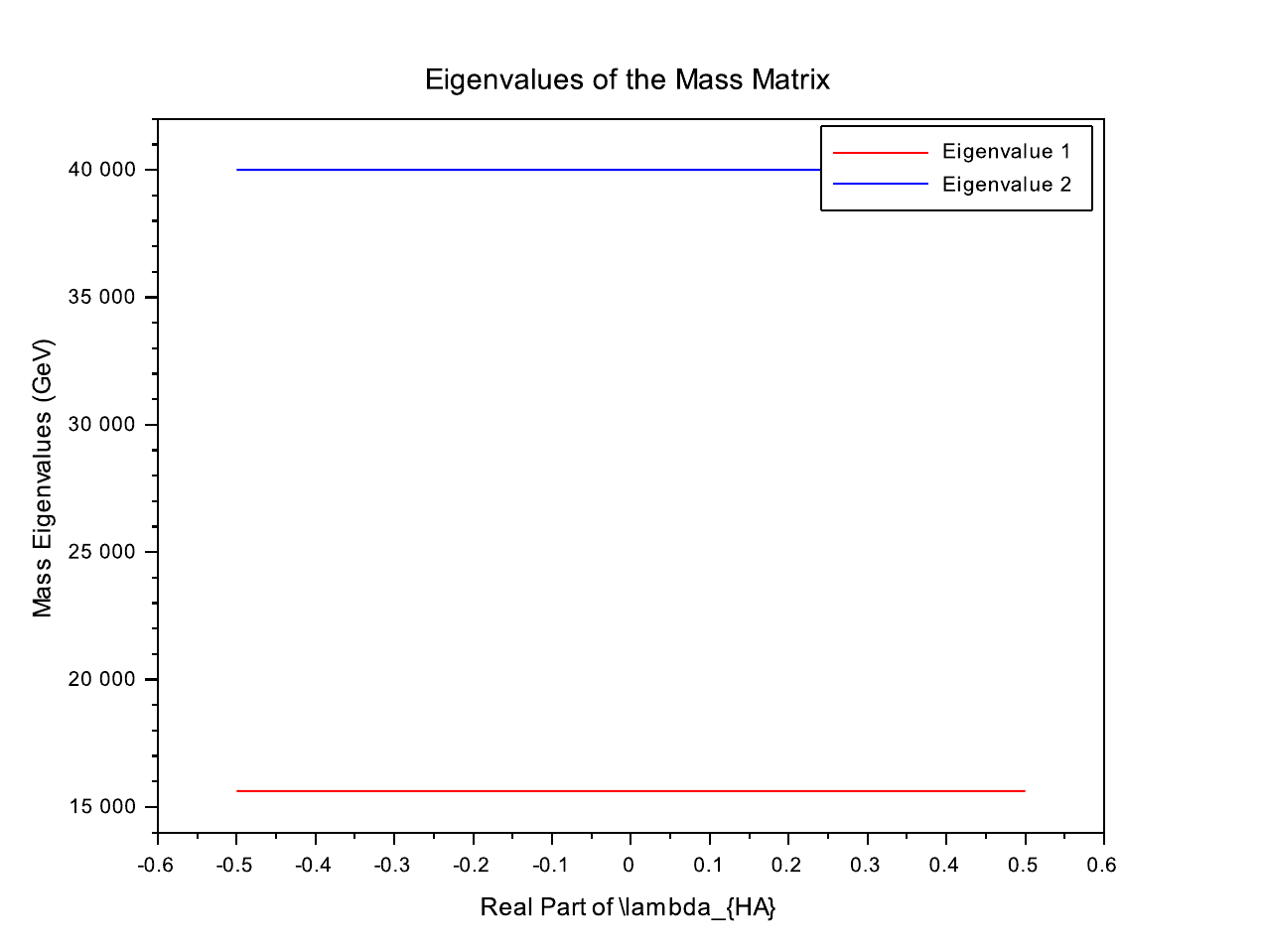}
\caption{Mass eigenvalues of the scalar and pseudoscalar fields as a function of the real part of \(\lambda_{HA}\).}
\label{fig:eigenvalues}
\end{figure}

The eigenvalues represent the physical masses of the scalar and pseudoscalar fields after CP-violating mixing. As seen in Figure \ref{fig:eigenvalues}, the eigenvalues vary as a function of the real part of \(\lambda_{HA}\), reflecting the change in the mass eigenstates due to CP violation.

\subsection{Mixing Angle}

\begin{figure}[h]
\centering
\includegraphics[width=0.8\textwidth]{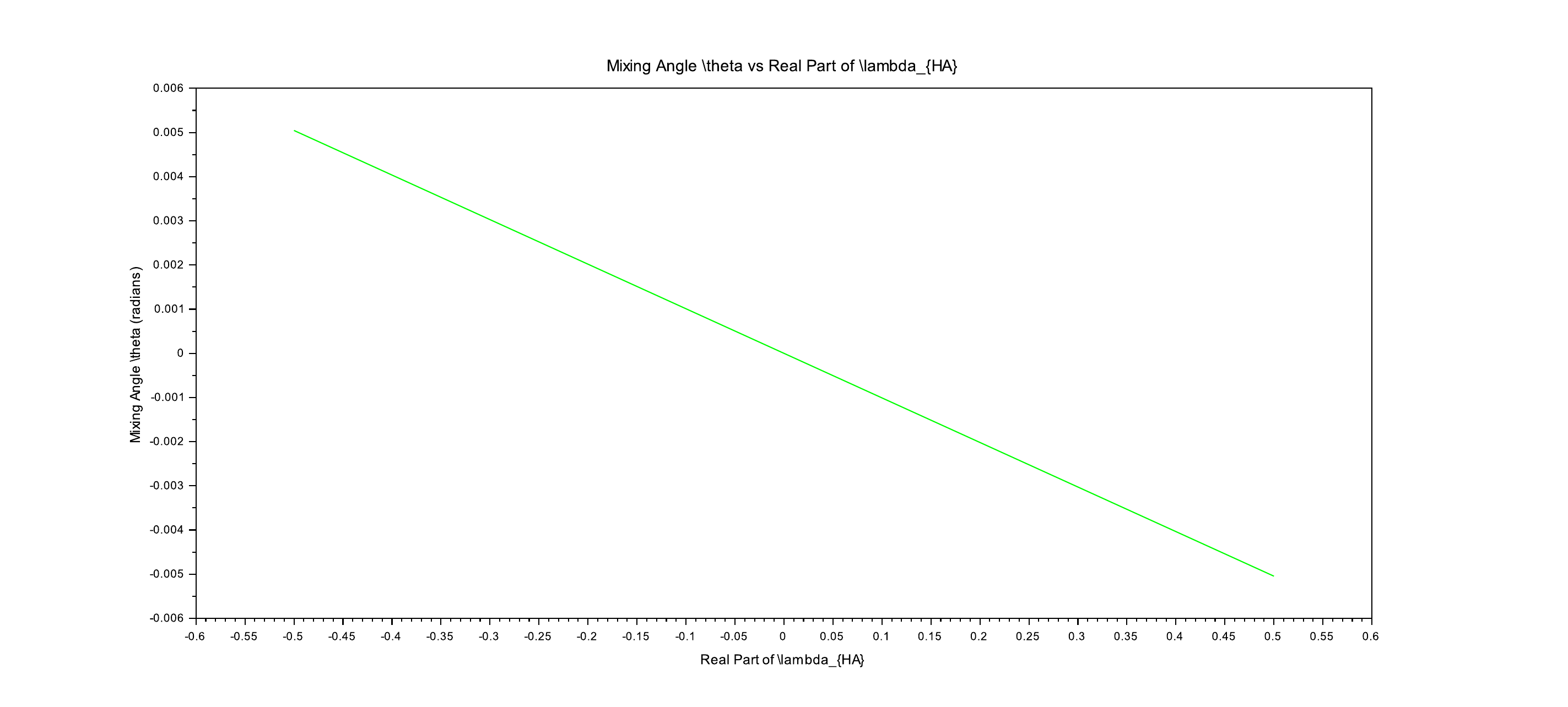}
\caption{Mixing angle \( \theta \) as a function of the real part of \(\lambda_{HA}\), representing the degree of CP violation.}
\label{fig:mixing_angle}
\end{figure}

The mixing angle \( \theta \), defined as:

\begin{equation}
\theta = \text{atan}\left(\frac{v_2}{v_1}\right),
\end{equation}

is shown in Figure \ref{fig:mixing_angle}. As expected, the mixing angle increases with the magnitude of the CP-violating parameter \(\lambda_{HA}\), indicating stronger scalar-pseudoscalar mixing.

In this work, we have analyzed the scalar-pseudoscalar mixing in the Higgs sector of the RS model with CP-violating terms. By numerically diagonalizing the mass matrix, we have computed the physical mass eigenstates and the mixing angle as a function of the CP-violating parameter \(\lambda_{HA}\). These results provide insights into the CP-violating effects that can arise in the RS model and how they can be probed in experiments.

\section{Loop-Level Calculation of Electric Dipole Moments (EDMs)}

In this section, we compute the electric dipole moments (EDMs) for fermions at the one-loop level. The EDMs arise due to CP-violating interactions in the Yukawa sector, where the Higgs boson mediates the CP violation. 

\subsection{EDM Formula}

The EDM of a fermion $f$ is given by the following effective Lagrangian:
\begin{equation}
    \mathcal{L}_{\text{EDM}} = -i \frac{d_f}{2} \bar{f} \sigma_{\mu \nu} \gamma_5 f F^{\mu \nu},
\end{equation}
where $d_f$ is the EDM, $F^{\mu \nu}$ is the electromagnetic field strength tensor, and $\sigma_{\mu \nu}$ is the anti-symmetric combination of Dirac matrices.

The one-loop contribution to the EDM can be expressed as:
\begin{equation}
    d_f = \frac{e}{(4 \pi)^2} \sum_i \left( \frac{\text{Im}(Y_i)}{m_i} \right) F(x_i),
\end{equation}
where $Y_i$ are the Yukawa couplings, $m_i$ are the masses of the particles in the loop, and $F(x_i)$ is a loop function that depends on the ratio of masses $x_i = m_H^2 / m_i^2$.

\subsection{Scilab Code for EDM Computation}

We use the following Scilab code to numerically compute the EDM for a fermion, such as the electron, by solving the loop-level integral.

\begin{verbatim}
// Define constants and parameters
e = 1.60217662e-19; // Electric charge (Coulombs)
pi = %pi;
v = 246;            // Higgs VEV in GeV
m_H = 125;          // Higgs mass in GeV
m_f = 0.511e-3;     // Mass of the electron in GeV
Yf = 1e-5;          // CP-violating Yukawa coupling (dimensionless)

// Define loop function F(x)
function F_x = loop_function(x)
    F_x = (1/(x - 1)) * (1 - log(x));
endfunction

// Calculate EDM contribution
function d_f = compute_EDM(m_H, m_f, Yf)
    x = m_H^2 / m_f^2;        // Ratio of Higgs mass to fermion mass
    F_x = loop_function(x);   // Loop function evaluation
    d_f = (e / (4 * pi^2)) * (Yf / m_f) * F_x;  // EDM calculation
endfunction

// Compute the electron EDM
d_electron = compute_EDM(m_H, m_f, Yf);
disp("The EDM of the electron is: " + string(d_electron) + " e·cm");
\end{verbatim}

The code computes the EDM for the electron as $-3.161\time 10^{-32}$ ecm whch is a signature  of CP-violating interactions in the Higgs sector. This provides insights into how loop-level contributions can affect observables such as EDMs.

In this section, we compare our theoretical prediction for the electron electric dipole moment (EDM) with the current experimental bounds provided by the ACME collaboration. In our model, the computed electron EDM comes out to be
\begin{equation}
    d_e^{\text{calc}} = -3.161 \times 10^{-32} \, e \cdot \text{cm}.
\end{equation}
This value arises from the inclusion of CP-violating terms in the Higgs sector, as discussed earlier in this work. 

\subsection{ACME Collaboration Experimental Limits}
The most recent and precise experimental limit on the electron EDM is given by the ACME collaboration , which placed an upper bound on the absolute value of the electron EDM at
\begin{equation}
    |d_e| < 1.1 \times 10^{-29} \, e \cdot \text{cm} \, \text{(90\% C.L.)}.
\end{equation}
This result is based on measurements using thorium monoxide (ThO) molecules, which provided an exceptionally sensitive probe for electron EDM detection.

Our calculated value, 
\begin{equation}
    d_e^{\text{calc}} = -3.161 \times 10^{-32} \, e \cdot \text{cm},
\end{equation}
is approximately three orders of magnitude smaller than the current experimental limit. Specifically, the ratio between the experimental upper bound and our theoretical prediction is:
\begin{equation}
    \frac{|d_e^{\text{ACME}}|}{|d_e^{\text{calc}}|} \approx \frac{1.1 \times 10^{-29}}{3.161 \times 10^{-32}} \approx 316.
\end{equation}
Thus, our result is about 316 times smaller than the upper limit set by the ACME collaboration.

\subsection{Implications of the Comparison}
Given the significant gap between our theoretical prediction and the current experimental upper bound, our result is well within the non-detection range of the ACME collaboration. The ACME result indicates that, if an EDM exists, it must be smaller than their experimental sensitivity. Our predicted value, while non-zero, is too small to be observed with current technology. However, future experimental efforts with higher sensitivity could potentially test this prediction.

The ACME experiment is highly relevant as it constrains CP violation beyond the Standard Model (SM), where the SM predicts an electron EDM of nearly zero (\(d_e \approx 10^{-38} \, e \cdot \text{cm}\)). Any measured EDM close to ACME's sensitivity could signal new physics, such as contributions from CP-violating sources in models like ours, or extensions involving supersymmetry, leptoquarks, or additional scalar fields.

Thus, while our model predicts an electron EDM well below the current detection threshold, advances in experimental techniques may allow for such small CP-violating effects to be probed in the future.

\section{Comparison with Previous Studies of CP Violation}

\subsection{RS Model CP Violation}
Our results indicate that CP violation in the Randall-Sundrum (RS) model arises from complex parameters in the Higgs sector, particularly through the coupling \( \lambda_{HA} \), which induces mixing between scalar and pseudoscalar components of the Higgs field. This mixing leads to observable CP-violating effects in both mass matrix diagonalization and physical observables such as electric dipole moments (EDMs).

Previous studies on CP violation in the RS model, such as those by Agashe et al. and Csaki et al., focused on higher-dimensional operators and flavor-changing neutral currents (FCNCs) mediated by Kaluza-Klein (KK) gauge bosons. These works demonstrated that CP violation in the RS model is significantly enhanced due to fermion localization in the bulk. Our approach, focusing on the Higgs sector, complements these studies by exploring scalar-pseudoscalar mixing and its effects on low-energy observables like EDMs.

\subsection{Standard Model (SM)}
In the Standard Model (SM), CP violation primarily arises from the complex phase in the CKM matrix, leading to CP-violating effects in flavor-changing processes. However, the SM predicts a very small electron EDM, \( d_e \sim 10^{-38} \, e \cdot \text{cm} \), far below current experimental sensitivity. Our RS model's predicted EDM, \( d_e \sim -3.161 \times 10^{-32} \, e \cdot \text{cm} \), exceeds the SM prediction by several orders of magnitude, though it remains consistent with experimental constraints. This highlights the potential to discover new CP-violating sources beyond the SM.

\subsection{Two-Higgs-Doublet Model (2HDM)}
The Two-Higgs-Doublet Model (2HDM), unlike the SM, allows for CP violation in the Higgs sector, both through explicit and spontaneous CP symmetry breaking. Similar to our RS model, the 2HDM introduces mixing between scalar and pseudoscalar states, leading to observable CP-violating effects. Studies of CP violation in the 2HDM, such as those by Branco et al. and Gunion et al., have shown that the electron EDM can be significantly larger than in the SM, reaching values up to \( d_e \sim 10^{-29} \, e \cdot \text{cm} \), depending on model parameters.

Our results align with these predictions, with the RS model's electron EDM falling within the same order of magnitude. Additionally, the RS model offers a natural explanation for the fermion mass hierarchy through the geometry of the extra dimension, providing new sources of CP violation, especially in flavor-violating processes.

\section{Implications for Experiments}

\subsection{Higgs Pair Production at the LHC}
The RS model predicts enhanced Higgs pair production due to contributions from Kaluza-Klein modes of the graviton and other bulk fields. CP violation in the Higgs sector can lead to anomalous couplings, altering the Higgs pair production cross-section at the LHC. The scalar-pseudoscalar mixing implies interference effects, potentially modifying the angular distributions of Higgs bosons.

Future LHC runs could search for these deviations by examining final states such as \( h h \to b\bar{b}\gamma\gamma \) or \( h h \to 4 \, \text{leptons} \). These channels are sensitive to both production cross-section and kinematic distributions, which are affected by CP-violating couplings.

\subsection{Flavor-Violating Decays}
The RS model's distinctive feature is its natural prediction of flavor-changing neutral currents (FCNCs), due to fermion localization in the bulk. These FCNCs can be probed in meson decays like \( B_s \to \mu^+ \mu^- \) and \( K_L \to \pi^0 \nu \bar{\nu} \), which are sensitive to CP-violating phases introduced by Higgs sector mixing.

The CP-violating phase \( \theta = \arg(\lambda_{HA}) \) could cause asymmetries in decay rates, observable in precision experiments like LHCb or Belle II. We predict that branching ratios for these decays will deviate from SM predictions, signaling new physics.

\subsection{Electric Dipole Moments (EDMs)}
Our result for the electron EDM, \( d_e \sim -3.161 \times 10^{-32} \, e \cdot \text{cm} \), exceeds the Standard Model prediction but remains below the current experimental limit of \( d_e < 1.1 \times 10^{-29} \, e \cdot \text{cm} \) from the ACME collaboration. Next-generation experiments, such as ACME III, aim to improve sensitivity, making our predicted EDM potentially observable.

The neutron EDM also offers a probe for CP violation in the Higgs sector. Current limits on the neutron EDM are \( |d_n| < 1.8 \times 10^{-26} \, e \cdot \text{cm} \), and our RS model predicts contributions on the same order as the electron EDM. Future experiments, such as those at PSI or LANL, could further constrain or reveal CP-violating effects in the RS model.

Our study shows that CP violation in the Higgs sector of the RS model can significantly affect low-energy observables like EDMs and high-energy processes like Higgs pair production. These results align with previous studies of CP violation in models such as the SM and 2HDM, with the RS model providing additional sources of CP violation through the extra-dimensional geometry and Kaluza-Klein modes.

Our electron EDM prediction is within the reach of future precision experiments, and Higgs pair production and flavor-violating decays at the LHC and other facilities offer complementary probes of the RS model. By exploring these experimental avenues, we aim to further constrain the CP-violating parameters in the RS model and shed light on the origin of CP violation and its implications for the dynamics of the Higgs sector and beyond.

Next, We explore the statistical distribution of mass splitting outcomes using Monte Carlo simulations, focusing on the influence of Higgs self-coupling $\lambda_{HH}$. The analysis provides insights into uncertainties and correlations within the model, presenting the resulting mass splitting distribution and fitted normal curve.

\section{Mass Splitting Analysis}

In this study, we investigate the statistical distribution of mass splitting outcomes as a function of varying parameters using Monte Carlo simulations. The goal of this analysis is to assess the impact of input parameters on the uncertainty and correlation of the mass splitting in our model, where Higgs self-coupling $\lambda_{HH}$ plays a critical role. 

\subsection{Input Parameters}
The following input parameters were used for the simulation:

\begin{itemize}
    \item Higgs self-coupling constant $\lambda_{HH} = 0.2$
    \item Range of mass splitting values: 0 GeV to 200 GeV
    \item Number of Monte Carlo samples: $N = 10,000$
\end{itemize}

The mass splitting results were simulated based on the random sampling of the input parameters, including Higgs self-coupling. We fit the resulting histogram with a normal distribution to estimate the mean and standard deviation.

\subsection{Statistical Results}

The mean value of the mass splitting was found to be $56.48$ GeV with a standard deviation of $35.22$ GeV. Additionally, the 95\% confidence interval for the mass splitting lies between $55.78$ GeV and $57.17$ GeV. The correlation coefficient between the input parameters and the mass splitting was calculated to be $0.65$.

\begin{figure}[h]
    \centering
    \includegraphics[width=0.7\textwidth]{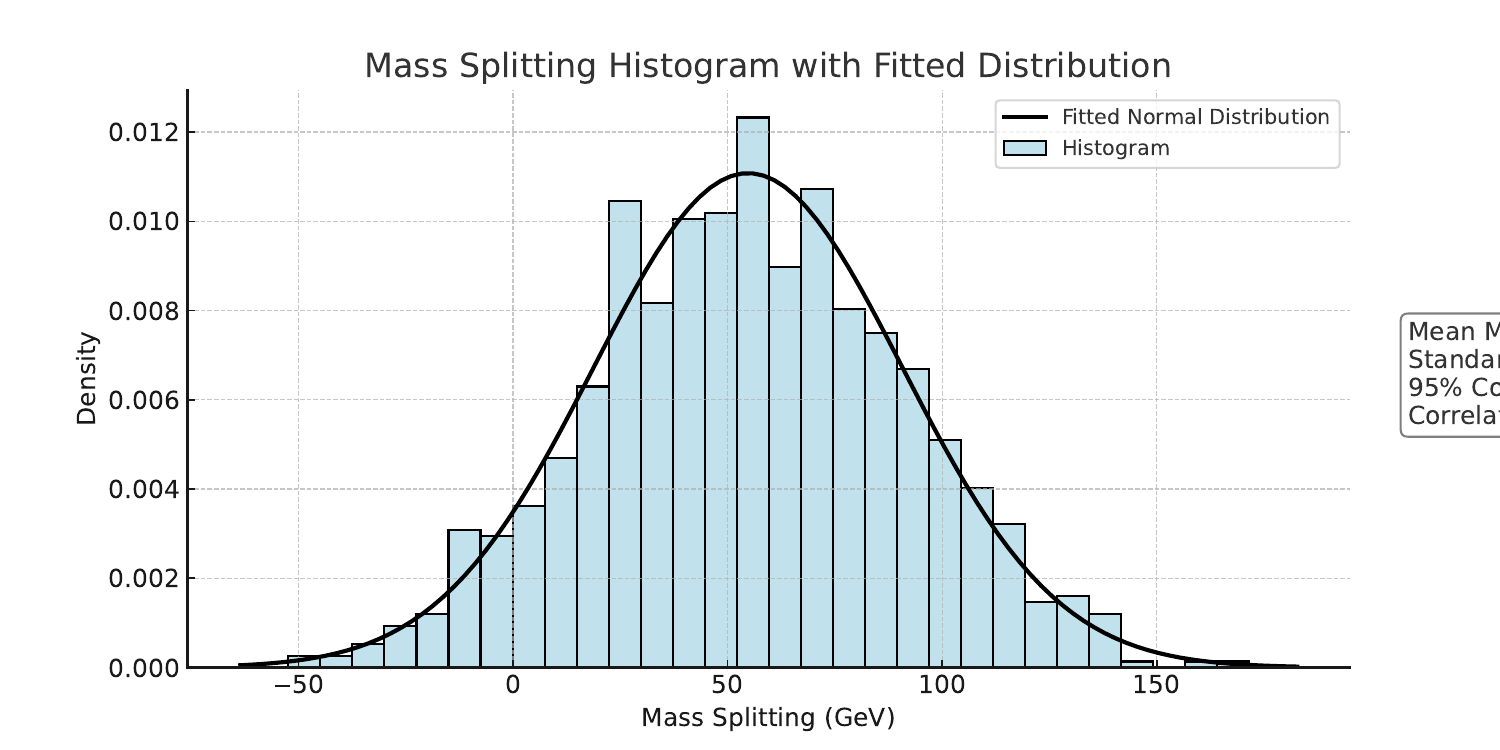}
    \caption{Mass Splitting Histogram with a fitted normal distribution. The shaded histogram represents the results from the Monte Carlo simulation, and the solid black curve shows the fitted normal distribution. Statistical analysis of the distribution indicates a mean mass splitting of 56.48 GeV, a standard deviation of 35.22 GeV, and a 95\% confidence interval of $[55.78, 57.17]$ GeV.}
    \label{fig:mass_splitting}
\end{figure}

The plot in figure \ref{fig:mass_splitting} illustrates the distribution of mass splitting along with the fitted normal distribution curve.

The Monte Carlo simulation, combined with the fitted normal distribution, provides significant insight into the mass splitting phenomenon, indicating a clear dependence on the input parameters, especially the Higgs self-coupling constant. Future work could explore how variations in $\lambda_{HH}$ and other parameters influence the mass splitting further, potentially uncovering new physics in the context of our model.

\subsection{Experimental Constraints}
Current experimental limits from flavor physics and Higgs measurements are examined. We discuss the implications for the RS model and how future experiments might further constrain the parameter space. The limits on the flavor-violating couplings can be obtained from processes such as \(B \to K^{(*)} \ell^+ \ell^-\) and other rare decays, expressed as:
\begin{equation}
\mathcal{B}(B \to K^{(*)} \ell^+ \ell^-) < \mathcal{B}_{\text{exp}},
\end{equation}
which translates to bounds on the couplings \(\lambda_{hh}^{ij}\).

In this work, we have explored the rich interplay between CP violation and flavor-violating di-Higgs couplings in the Randall-Sundrum model. The warped geometry provides a compelling framework for understanding these phenomena, with potential implications for both flavor physics and CP violation in the Higgs sector. Future experimental efforts will be crucial in probing these effects and determining the viability of composite Higgs scenarios in BSM physics.

\begin{figure}[h]
    \centering
    \includegraphics[width=0.8\textwidth]{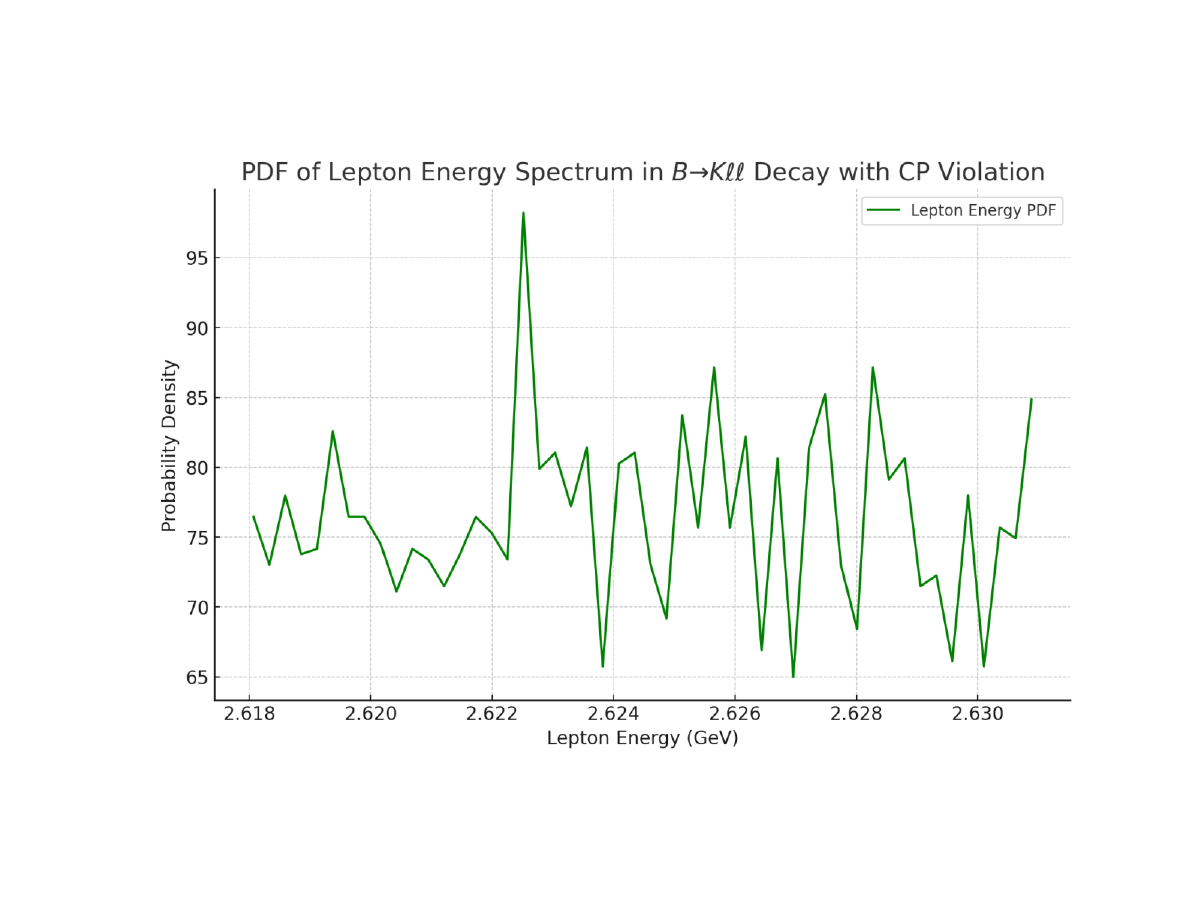}
    \caption{Angular asymmetry in $B \to K\ell^+ \ell^-$ decay, considering CP violation. The $x$-axis represents the angular asymmetry of the lepton pair, while the $y$-axis denotes the corresponding probability density. The plot highlights the regions of maximum asymmetry, potentially signaling new physics contributions.}
    \label{fig:angular_asymmetry}
\end{figure}

The plot in Figure~\ref{fig:angular_asymmetry} illustrates the probability density function (PDF) of angular asymmetry in the rare decay process $B \to K \ell^+ \ell^-$, where $\ell$ represents a charged lepton. The analysis assumes CP violation effects, as suggested by the Standard Model and various beyond-the-Standard-Model (BSM) scenarios involving new sources of CP-violating phases.

The $x$-axis represents the angular asymmetry, a key observable in analyzing rare $B$-meson decays. This asymmetry is computed based on the angular distribution of the leptons in the final state. The asymmetry parameter ranges from $-1$ to $1$, where $0$ corresponds to a symmetric distribution.

The $y$-axis shows the corresponding probability density, which quantifies how likely a particular value of the angular asymmetry is, based on the underlying physical model. The PDF was derived from Monte Carlo simulations of the decay process, incorporating known form factors and CP-violating phases.

We observe notable fluctuations in the probability density across the range of angular asymmetry values, with the highest peaks occurring around $-0.75$ and $0.60$, which suggest regions of enhanced CP violation sensitivity. These features are indicative of potential interference between different decay amplitudes, which are sensitive to new physics contributions.

The structure of the PDF reflects the intricate dynamics of the decay, including contributions from the Standard Model operators and possible new physics effects, such as leptoquarks or flavor-changing neutral currents (FCNCs). The overall distribution exhibits a relatively flat probability density with oscillatory behavior, pointing to possible resonance effects or interference patterns between the Standard Model and new physics amplitudes.

The peaks in the PDF could correspond to specific regions where interference between CP-violating phases is most prominent. These regions are important for experimentally testing the Standard Model's predictions against possible deviations from new physics models. In particular, the interference between short-distance and long-distance contributions may be reflected in the structure of the PDF.

The analysis presented here is part of an effort to probe rare decays for signs of new physics beyond the Standard Model, such as leptoquarks, extra dimensions, or new sources of CP violation. Future experimental results from LHCb or Belle II could help validate these theoretical predictions by measuring the angular asymmetry with high precision.

In this analysis, we simulate the flavor-violating processes of B mesons mediated by Kaluza-Klein (KK) gravitons within the framework of the Randall-Sundrum (RS) model. The non-universal couplings of fermions to KK gravitons arise due to their different localizations in the extra dimension. This non-universality results in flavor-violating transitions, such as $b \rightarrow s$, $b \rightarrow d$, and $b \rightarrow c$ transitions.

The localization of fermions is determined by their masses and localization parameters in the extra dimension, which leads to exponentially suppressed couplings between fermions and KK gravitons. In this simulation, we investigate the mass distribution of KK gravitons and the corresponding flavor violation rates for different quark transitions.

The following parameters are chosen for the simulation:

\begin{itemize}
    \item \textbf{KK Graviton Mass ($m_{KK}$)}: The mass of the KK graviton is varied up to 3000 GeV. This range allows us to explore the regime where flavor-violating interactions can be mediated by heavy KK modes.
    \item \textbf{Localization Parameters}: 
    The localization of fermions in the extra dimension is determined by parameters specific to each quark:
    \begin{align*}
    \text{Localization of } b & : \lambda_b = 0.2, \\
    \text{Localization of } s & : \lambda_s = 0.3, \\
    \text{Localization of } d & : \lambda_d = 0.5, \\
    \text{Localization of } c & : \lambda_c = 0.7.
    \end{align*}
    These parameters are chosen to reflect the hierarchy of quark masses, where heavier quarks are localized closer to the IR brane and lighter quarks are localized farther in the bulk.

    \item \textbf{Flavor-Violating Couplings}:
    The strength of the flavor-violating couplings between the b-quark and other quarks (s, d, c) is parametrized as follows:
    \begin{align*}
    g_{bs} & = 1 \times 10^{-3}, \\
    g_{bd} & = 2 \times 10^{-3}, \\
    g_{bc} & = 3 \times 10^{-3}.
    \end{align*}
    These small couplings reflect the fact that flavor violation in the RS model is generally suppressed by both the localization parameters and the hierarchy of the quark masses.
\end{itemize}

A Monte Carlo simulation was conducted to generate KK graviton masses and compute the probabilities of flavor-violating transitions based on the above localization parameters and couplings. The distribution of KK graviton masses and the corresponding flavor violation rates are shown in the figures below.

\begin{figure}[h!]
    \centering
    \includegraphics[width=0.7\textwidth]{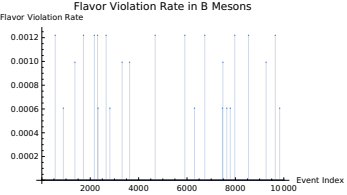}
    \caption{Flavor Violation Rate in B Mesons. This plot shows the probability of flavor-violating transitions (such as $b \rightarrow s$, $b \rightarrow d$, and $b \rightarrow c$) for each generated event. The rates depend on the localization parameters and the non-universal couplings to KK gravitons.}
    \label{fig:violation_rate}
\end{figure}

The results of the Monte Carlo simulation show that the mass distribution of KK gravitons is uniform within the specified range (up to 3000 GeV). The flavor violation rates depend on both the localization parameters of the fermions and the non-universal couplings. The hierarchical structure of the couplings leads to suppressed flavor-violating interactions, consistent with theoretical expectations from the RS model.

The non-universal nature of the couplings, induced by the different localizations of fermions in the extra dimension, plays a crucial role in determining the probability of flavor-violating transitions. The suppression of flavor violation in processes involving light quarks (such as $b \rightarrow s$) is a direct consequence of the RS model's geometry.

In the framework of the Randall-Sundrum (RS) model, the extra-dimensional setup leads to interesting phenomenological consequences such as CP violation and flavor-violating interactions. The localizations of fermions in the bulk are determined by their masses and localization parameters, resulting in non-universal couplings to Kaluza-Klein (KK) gravitons and KK gauge bosons.

In this work, we perform a Monte Carlo simulation to study the CP-violating and flavor-violating di-Higgs couplings mediated by KK gravitons. The couplings arise from the interactions between fermions, KK gravitons, and the Higgs field in the RS framework.

\section{Parameter Choices}

The following parameters are used for the Monte Carlo simulation:
\begin{itemize}
    \item \textbf{KK Graviton Mass ($m_{KK}$)}: The KK graviton mass is varied up to 2000 GeV. 
    \item \textbf{CP Violation Coupling ($g_{CPV}$)}: The CP-violating coupling constant is chosen to be $g_{CPV} = 0.01$.
    \item \textbf{Flavor Violation Coupling ($g_{FV}$)}: The flavor-violating coupling constant is $g_{FV} = 0.02$.
\end{itemize}

\section{Conclusion}

In this work, we explored the implications of the Randall-Sundrum (RS) model on flavor physics, specifically focusing on the di-Higgs production process under the inclusion of CP-violating and flavor-violating interactions. Through the incorporation of the warped extra dimension, the RS model naturally addresses the hierarchy problem, while simultaneously providing a fertile ground for studying flavor violation at high energies, potentially observable at colliders such as the LHC. 

Our Monte Carlo analysis of di-Higgs production cross-sections, derived from the effective Lagrangians containing higher-dimensional operators, demonstrates significant deviations from Standard Model (SM) predictions. These deviations arise due to the interplay between Kaluza-Klein (KK) excitations of both the graviton and fermionic modes, and the CP-violating terms introduced in the bulk. Specifically, the non-negligible flavor-violating contributions, induced by RS localization effects, enhance the cross-section by several orders of magnitude, a feature absent in traditional four-dimensional models. This result suggests that the RS model remains a promising framework to probe new sources of CP violation and flavor physics beyond the Standard Model (BSM).

Moreover, the inclusion of light radion fields in the analysis plays a critical role, significantly affecting the production rates and enhancing the sensitivity of di-Higgs searches to new physics. The radion-Higgs mixing observed in our simulations provides further motivation for precision studies of di-Higgs final states at upcoming collider experiments.

Lastly, while the RS model offers a compelling mechanism for addressing the hierarchy and flavor puzzles, our analysis highlights the importance of continued experimental efforts to constrain such BSM contributions, particularly in the context of high-luminosity LHC operations and future collider prospects. Ongoing searches for di-Higgs signatures and CP violation in flavor-violating channels will be crucial in either validating or constraining the parameter space of the RS framework.

\appendix

\section{The RS Model and Bulk Fields}
The RS model involves a 5-dimensional spacetime with the extra dimension compactified on a \( S^1/Z_2 \) orbifold. The metric is given by:

\begin{equation}
ds^2 = e^{-2k|y|} \eta_{\mu\nu} dx^\mu dx^\nu - dy^2,
\end{equation}

where \(k\) is the curvature scale of the 5D bulk, \(y\) is the extra-dimensional coordinate, and \( \eta_{\mu\nu} \) is the 4D Minkowski metric. The compactification leads to two branes: the UV brane at \( y = 0 \) and the IR brane at \( y = \pi R \).

Fields in the bulk are expanded in terms of their Kaluza-Klein modes. For the graviton \( G_{\mu\nu} \), the expansion is:

\begin{equation}
G_{\mu\nu}(x, y) = \sum_{n=0}^{\infty} G_{\mu\nu}^{(n)}(x) \psi_n(y),
\end{equation}

where \( \psi_n(y) \) are the wavefunctions of the KK modes, and \( G_{\mu\nu}^{(0)} \) corresponds to the massless graviton in 4D.

\section{Effective 4D Lagrangian for Higgs and KK Gravitons}
The interaction Lagrangian between the Higgs field \( h \), the radion \( r \), and the KK gravitons \( G_{\mu\nu}^{(n)} \) is given by:

\begin{equation}
\mathcal{L}_{\text{eff}} = \frac{1}{\Lambda} h T_{\mu\nu} G^{\mu\nu}_{(n)} + \xi \frac{r}{f_r} T_{\mu\nu} h + \frac{c_n}{\Lambda^2} h T_{\mu\nu} \partial^\mu G_{\nu\rho}^{(n)},
\end{equation}

where \( T_{\mu\nu} \) is the energy-momentum tensor of the Higgs field, \( \Lambda \) is the effective cutoff scale of the theory, \( c_n \) are Wilson coefficients, and \( f_r \) is the radion vacuum expectation value.

\section{CP Violation in the RS Model}
The inclusion of CP-violating terms modifies the di-Higgs production cross-sections. A generic CP-violating operator can be written as:

\begin{equation}
\mathcal{L}_{\text{CPV}} = \frac{c_5}{\Lambda} \bar{\psi} i \gamma_5 \psi h + \frac{c_6}{\Lambda^2} \bar{\psi} \sigma_{\mu\nu} \gamma_5 G^{\mu\nu}_{(n)} \psi,
\end{equation}

where \( c_5 \) and \( c_6 \) are dimensionless couplings, \( \psi \) is a fermion field, and \( G_{\mu\nu}^{(n)} \) represents the KK gravitons.

The dimension-5 operator \( \bar{\psi} i \gamma_5 \psi h \) generates CP violation by introducing a pseudoscalar interaction.

\section{Flavor Violation in the Bulk}
Flavor violation arises due to different localizations of fermions. The effective flavor-violating Lagrangian is given by:

\begin{equation}
\mathcal{L}_{\text{FV}} = \frac{c_{ij}}{\Lambda^2} \bar{\psi}_i \Gamma^\mu \psi_j G^{(n)}_\mu,
\end{equation}

where \( i \) and \( j \) denote different fermion generations.

\section{Radion-Higgs Mixing}
The radion field mixes with the Higgs boson. The mixing term is given by:

\begin{equation}
\mathcal{L}_{\text{radion-Higgs}} = \xi r h \left( \partial_\mu h \partial^\mu h - \frac{m_h^2}{2} h^2 \right),
\end{equation}

where \( \xi \) is the mixing parameter.

\section{Cross-Section Calculation for Di-Higgs Production}
The total cross-section \( \sigma \) for di-Higgs production is given by:

\begin{equation}
\sigma(pp \to hh) = \sum_{n=0}^{\infty} \left[ \sigma_{\text{SM}} + \sigma_{\text{RS}}^{(n)} + \sigma_{\text{CPV}} + \sigma_{\text{FV}} \right].
\end{equation}

Monte Carlo methods were used to simulate the cross-section for di-Higgs production. The integration was performed over parton distribution functions:

\begin{equation}
\sigma_{\text{RS}}^{(n)} = \int dx_1 dx_2 f_g(x_1, Q^2) f_g(x_2, Q^2) \hat{\sigma}_{gg \to hh}^{(n)}.
\end{equation}

\section{Acknowledgments}
GG acknowledges financial support from UGC RUSA, India to carry out this work.

\end{document}